\documentclass[usenatbib]{mnras}
\usepackage{times,graphics,amssymb,epsfig,subfigure,color}

\usepackage[T1]{fontenc}
\usepackage{ae,aecompl}
\usepackage{float}
\usepackage{lipsum}
\usepackage{graphicx}
\usepackage{cleveref}
\usepackage{dblfloatfix}
\usepackage{float}

\newcommand{\ber}{\begin{eqnarray}}
\newcommand{\eer}{\end{eqnarray}}

\def\labfig #1{\label{fig:#1}}
\def\labsecn #1{\label{sec:#1}}
\def\labsubsecn #1{\label{subsecn:#1}}

\def\fig #1{Figure~\ref{fig:#1}}
\def\secn #1{Section~\ref{sec:#1}}

\def\subsecn #1{Section~\ref{subsecn:#1}}

\def\dsubsecn #1#2{Sections~{\ref{subsecn:#1}}~and~{\ref{subsecn:#2}}}

\def\etal{et al.\ }

\def\unit #1{\,{\rm #1}}

\def\ev{\unit{eV}}
\def\kev{\unit{keV}}

\title[Faint BEL and NEL for HMBH]
{Hyper-massive Black Holes have Faint Broad and Narrow Emission Lines}

\author[H. K. Bhat \etal]
{Harshitha K. Bhat$^{1,2,3}$
\thanks{E-mail: harshikikkeri@gmail.com (HB)},
Susmita Chakravorty$^{4}$
\thanks{Email: write2susmita@gmail.com (SC)}, 
Dhrubojyoti Sengupta$^{5}$, \and 
Martin Elvis$^{6}$, 
Sudeb Ranjan Datta$^{4}$,
Nirupam Roy$^{4}$,
Caroline Bertemes$^{7}$, \and 
Gary Ferland$^{8}$, 
Savithri H. Ezhikode$^{9}$\\
\\
\footnotesize \it $^{1}$Department of Physics, St. Joseph's College, 36 Lalbagh Road, Bangalore 560027, India \\
\it $^{2}$I.Physikalisches Institut der Universit\"at zu K\"oln, Z\"ulpicher Str. 77, 50937 K\"oln, Germany \\
\it $^{3}$Max-Plank-Institut f\"ur Radioastronomie, Auf dem H\"ugel 69, 53121 Bonn, Germany \\
\it $^{4}$Department of Physics, Indian Institute of Science, Bangalore 560012, India; write2susmita@gmail.com \\
\it $^{5}$Presidency University, 86/1 College Street, Kolkata 700073, India\\
\it $^{6}$Harvard-Smithsonian Center for Astrophysics, 60 Garden Street, Cambridge, MA 02138, USA \\
\it $^{7}$ Department of Physics, University of Bath, Claverton Down, Bath BA2 7AY, UK\\
\it $^{8}$Department of Physics and Astronomy, University of Kentucky, Lexington, KY 40506, USA \\
\it $^{9}$Inter-University Centre for Astronomy \& Astrophysics (IUCAA), Ganeshkhind, Pune, India 
}
\begin{document}

\maketitle


\begin{abstract}
	The EUV provides most of the ionization that creates the high
	equivalent width (EW) broad and narrow emission lines (BELs, NELs) of
	quasars. Spectra of Hypermassive Schwarzschild black holes (HMBHs,
	$M_{BH} \geq 10^{10} M_{\odot}$) with $\alpha$-discs, decline rapidly
	in the EUV suggesting much lower EWs. Model spectra for black holes of
	mass $10^{6}-10^{12} M_{\odot}$ and accretion rates $0.03 \leq
	L_{bol}/L_{edd} \leq 1.0$ were input to the CLOUDY photoionization
	code. BELs become $\sim$100 times weaker in EW from $M_{BH} \sim 10^8
	M_{\odot}$ to $M_{BH} \sim 10^{10} M_{\odot}$. The high ionization BELs
	(O VI 1034 \AA $ $, C IV 1549 \AA $ $, He II 1640 \AA) decline in EW
	from ($M_{BH} \geq 10^6 M_{\odot}$, reproducing the Baldwin effect, but
	regain EW for $M_{BH} \geq 10^{10} M_{\odot}$). The low ionization lines
	(MgII 2798 \AA, H$\beta$ 4861 \AA $ $ and H$\alpha$ 6563 \AA $ $)
	remain weak. Lines for maximally spinning HMBHs behave similarly. Line
	ratio diagrams for the BELs show that high OVI/H$\beta$ and low
	CIV/H$\alpha$ may pick out HMBH, although OVI is often hard to observe.
	In NEL BPT diagrams HMBHs lie among star forming regions, except for
	highly spinning, high accretion rate HMBHs. In summary, the BELs
	expected from HMBHs would be hard to detect using the current optical
	facilities. From 100 to $10^{12} M_{\odot}$, the emission lines used to
	detect AGN only have high EW in the $10^6 - 10^9 M_{\odot}$ window,
	where most AGN are found. This selection effect may be distorting
	reported distributions of $M_{BH}$.

\end{abstract}




\section{Introduction}
\label{sec:Introduction}    

Active galactic nuclei (AGN) are the central regions of galaxies that host Super-massive black holes (SMBHs) which are actively accreting surrounding material. They have been an exciting field of research in Astrophysics for along time now \citep[][and references therein]{Netzer15}. The gravitational potential energy of the infalling matter is converted into radiation which photoionizes both the unresolvable nearby ($\lesssim 0.1$ pc, to the black hole) and more distant ($\lesssim 100$ pc, partially resolvable in near AGN) gas which then re-emit as Broad and Narrow Emission Lines (BELs \& NELs) respectively, in the optical and UV regions. The NELs usually have Doppler widths $\leq 500$ km s$^{-1}$ and arise in relatively low density ($\sim 10^{3}$ cm$^{-3}$) gas. BELs have Doppler widths $\sim 1,000-10,000$ km s$^{-1}$ and arise in higher density gas ($\sim 10^{9}$ cm$^{-3}$) as determined by absence of certain forbidden lines. Such large Doppler widths in the BELs suggest that the broad line regions (BLRs) are in deep gravitational potential where Keplerian velocities are often > 1\% c, which makes BELs very important in understanding the central source. Furthermore, the presence of BELs is an indicator of AGN activity. 

Studying the BELs \& NELs in AGN and using them to understand fundamental properties, particularly the mass, of the associated black holes (BHs) is a well-established method.  New data from more and more sensitive multi-wavelength campaigns suggest the presence of Hyper-massive Black Holes (HMBHs) (M$_{BH}$ $\geq 10^{10}M_{\odot}$) which are extreme and/or unusual for AGN \citep[e.g. []{Fan01,Mortlock11,Wu15}. \cite{King15} gives a maximum theoretical limit for $M_{BH}$ through luminous accretion as $\sim 5 \times 10^{10} M_{\odot}$ for typical parameters. Of the $> 10^{5}$ quasars known only a few tens have black hole masses $>10^{10}M_{\odot}$ \citep{Natarajan09,Ichikawa17}.  This scarcity may be because they are rare, or perhaps cannot grow any larger.  Discovery of more HMBH will put serious constraints on the current models for seed black holes. \citep{Volonteri12, Johnson13, Natarajan14, Latif16}. Finding even one black hole with $M_{BH} \geq
10^{11}M_{\odot}$ at high redshift (z>6) would pose serious challenges for the existing black hole evolution theories, requiring either highly super-Eddington accretion or supermassive seed black holes. 

Here we investigate whether this paucity of HMBH could be a selection effect due to the cooler temperatures of alpha-disc accretion discs at high masses. We were motivated by the striking results for low mass ($< 10^5 M_{\odot}$) BHs by \citet{Chakravorty14}, who found that BLR emission is drastically (factor $> 10$ for H$\beta$ 4861 \AA $ $) and quite suddenly (in $< 1$ dex in mass from $10^5$ to $10^4$ $M_{\odot}$) reduced as the discs become too hot and overionize gas moving at Keplerian speeds $> 1000$ km/s. A complementary effect is expected as the discs become too cool (in case of HMBHs) to photoionize the gas to produce the typical BLR emission lines. That cooler accretion discs would result in weaker line emission, was first investigated analytically by \citep{Laor11}. Hence we can expect that the emission lines will be weaker for HMBHs.  Here we make a detailed investigation of the predicted BLR and NLR equivalent widths (EWs) as a function of HMBH mass.

As the mass estimates for AGN come from the detection of broad emission lines in the UV and optical region of the  AGN spectra, we wanted to know if the same methods of detection are possible for HMBH. Therefore, we investigate the predicted EWs of few prominent BELs over a wide range of black hole masses ($M_{BH}=10^{6}-10^{12}M_{\odot}$) and accretion rates ($L_{bol}/L_{edd}=0.03-1$), to see if there is a decrease in intrinsic line strengths among these BELs, making such BHs very hard to be discovered with our current standard detection methods. See section \ref{sec:Broad Emission Lines}.

The strengths of the BELs and NELs depend on the ionizing spectral energy distribution (SED), which in turn depends on the fundamental properties of the BH: its mass and the rate at which it is accreting matter. Larger BHs have cooler maximum temperatures. At some mass the ionizing photons they emit will drop, potentially weakening their BELs and NELs to the point of undetectability. Investigating whether this happens for HMBHs is the purpose of this paper. The mass accretion rate ($\dot{m}$) can be related to the ratio between the bolometric luminosity ($L_{bol}$) of the accretion disc of accreted matter around the black hole, to the classical Eddington luminosity $L_{edd}$ of the BH. In Section \ref{sec:Spectral Energy Distribution} we shall show the details of how we have related the accretion rate with $L_{bol/L_{Eddd}}$. Thus, throughout the paper, the accretion rate will be represented by $L_{bol}/L_{edd}$.

Since we are interested in the higher end of the BH masses, we could not rely on the conventional method of assuming broken power-laws to construct the broad band SEDs of the BHs \citep{Tananbaum79, Lusso16}. In section \ref{sec:Spectral Energy Distribution}, we elaborate on the methods adopted to properly link the different components of the AGN SED, while being careful about the mass evolution of these inter-relations. 

Unlike BELs, NELs are not necessarily signatures of AGN activity. Dynamically (velocity widths), NELs are very similar to the emission lines of star-forming regions, planetary nebulae and even supernova remnants, as the photoionization equilibrium in all these regions is achieved at around same temperatures (T$\sim 10^{4}$ K). To distinguish between narrow emission lines from all these different astrophysical objects, BPT diagrams \citep{BaldwinPhillipsTerlevich81} were made as an attempt to have diagnostic diagrams for NELs in SMBH. Using our predicted strengths of NELs we also investigate the line ratios of HMBHs on BPT diagrams. See section \ref{sec:Narrow Emission Lines}. Since the BPT diagrams are well-tested tools to find NELs, in AGN, we extend the same logic to look at BELs. Since now we are looking at a much wider range of physical parameters of the black holes, we do not expect the broad lines to show intensity variations within a standard small range, anymore. See section \ref{subsec:BelLR} for the interesting consequences of mass and accretion rate variation on the line ratios of different BELs.  

In this study, we have used the photoionization code called CLOUDY \footnote{{http://www.nublado.org/}} \citep{Ferland17}, to predict the emission line strengths as a function of $M_{BH}$ and $L_{bol}/L_{edd}$. The physical parameters used for the CLOUDY calculations are mentioned in detail in Section \ref{sec:CLOUDY Calculation}. To model the gas/clouds that produce the BELs and NELs, we use the ``locally optimally emitting cloud (LOC)'' model \citep{Baldwin95}, which suggests that the emission lines are produced optimally - that each line is produced most efficiently within a narrow range of density and incident light flux. The history of this model and how we use it in this paper have been detailed in Section \ref{sec:LOC Model}.

\section{Spectral Energy Distribution}
\label{sec:Spectral Energy Distribution}

\subsection{The radiation from the accretion disc}
\labsubsecn{subsec:Diskbb}

The main source of energy in AGN is the conversion of gravitational potential energy of the accreting matter into radiation. This radiation from the accretion disc around the black hole, peaks in the extreme-ultraviolet region (EUV, 10-100eV). The SED of AGN is broad band, extending from radio wavelengths to gamma rays. However, for the study of photoionized gas, the range from optical-UV to soft X-rays matters most, because these photons ionize the gas present in the central $\sim$100 pc of the AGN. Observationally there is uncertainty in the shape of the SED of the radiation from the accretion disc near its peak, due to absorption of EUV radiation by Galactic neutral gas and dust. Therefore, observations below 10 eV and above 100 eV have to be used to reconstruct the unobserved part of SEDs.

According to standard theory of accretion discs \citep{Shakura73, Frank02}, radiation from the accretion disc can be modeled as the sum of all the local blackbodies emitted by different annuli of the disc. By knowing the temperature of the innermost stable annulus of the disc, temperature of all the subsequent rings can be estimated. A simplified form of the temperature of the annulus at radius $R$ is given by Equation 3.19 of \citet{Peterson97} (also see the `large r' approximation of Equation 5.43 of \citet{Frank02}). The same equation can be rewritten as: 
\begin{equation} 
\label{eq:Temperature}
T(R) = 3.5 \times 10^5 \left( \eta \right)^{-\frac{1}{4}} \left(\frac{\dot{m}} {\dot{m}_{\rm Edd}} \right)^{\frac{1}{4}} \left(\frac {M_{BH}} {10^8M_{\odot}} \right)^{-\frac{1}{4}} \left(\frac{R}{R_s}\right)^{-\frac{3}{4}}\rm{K} 
\end{equation}
$M_{BH}$ is the mass of the central black hole, $\dot{m}$ is its mass acrretion rate, while $\dot{m}_{\rm Edd}$ is the Eddington mass accretion rate. $R_s =
2GM_{BH}/c^2$ is the Schwarzschild radius, $G$ being the gravitational constant and $c$ the velocity of light. $\eta$ is the accretion efficiency factor, which
relates \begin{equation}
\label{eq:L_eta}
L = \eta \dot{m} c^2
\end{equation}
$L$ being the luminosity emitted from the disk (both surfaces, top and bottom). 

A standard model for the spectral component of a non-spinning accretion disc is available as a package called $diskbb$ in XSPEC \footnote{{https://heasarc.gsfc.nasa.gov/docs/xanadu/xspec/}}\citep{Mitsuda84, Makishima86}. The two inputs required by $diskbb$ are $T(R_{in})$ (which can be derived from \ref{eq:Temperature}) and the normalisation $A_{dbb}$, given by,
\begin{equation}
\label{eq:Adbb}
	A_{dbb} = \left\{ \frac {R_{in}/\rm{km}} {D/(10\,\,\rm{kpc})} \right\}^2 \cos\theta
\end{equation}
for a black hole observed at a distance $D$ (= 100 Mpc throughout this paper) whose line-of-sight makes an angle $\theta$ (=$30^{\circ}$ throughout this paper, unless otherwise mentioned) to the normal to the plane of the disc. $R_{in}$ is the distance of the innermost stable annulus of the accretion disc from the central black hole. 

\begin{figure*}
\begin{center}
\includegraphics[width=0.9\textwidth, trim= 0 10 0 0]{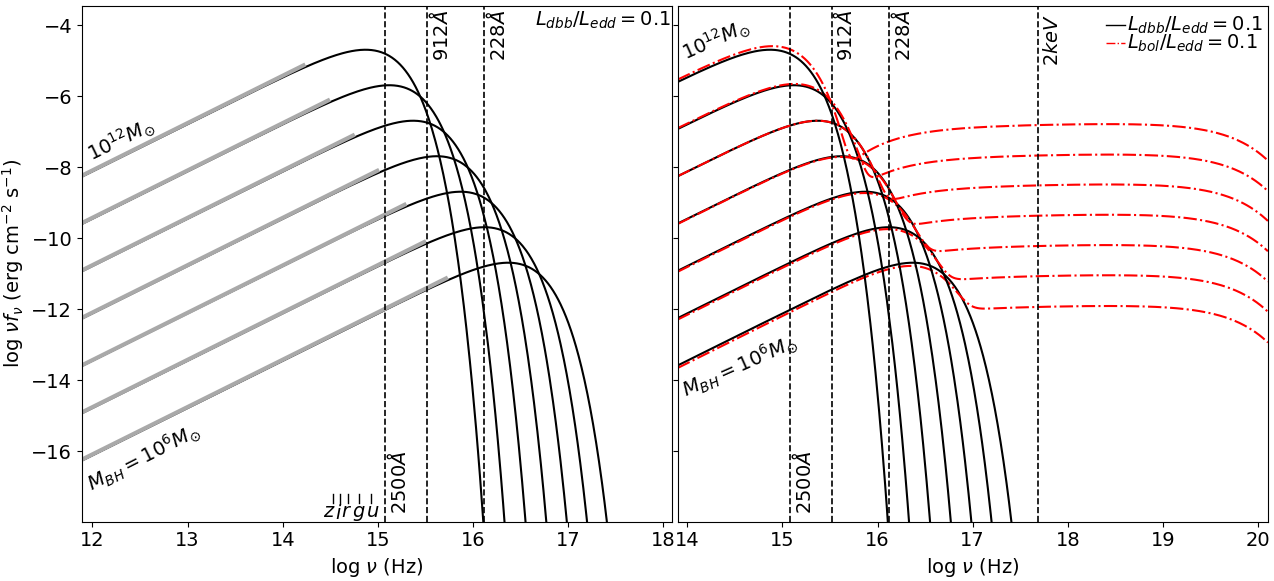}
	\caption{{\it Left Panel}: The spectral energy distributions (SEDs) of
the radiation from the accretion disc around the super-to-hyper-massive black
holes. The SEDs are calculated using the package $diskbb$ within XSPEC. All the
black holes here, are accreting at the rate $L_{dbb}/L_{edd} = 0.1$ and the
different lines correspond to different black hole masses from $M_{BH} = 10^6 -
10^{12} M_{\odot}$ at steps of 1 dex. The gray thicker lines highlight the
region of the SED where $f_{\nu} \sim \nu^{\frac{1}{3}}$. Note that the
traditionally (since Tananbaum et al. (1979)) important wavelength $2500$\AA $
$ does not fall in the gray part of the SED, unless $M_{BH} \lesssim 10^8
M_{\odot}$. We have also marked the positions of the centres of the five
filters of SDSS, namely u, g, r, i and z. {\it Right panel}: The complete AGN
SED from optical to X-rays (red dotted-and-dashed lines), but consisting of
only the components from the central regions - the $diskbb$ component, and the
power-law component. The procedure of appropriately adding these two components
to get the complete SED is elaborately described in Section \ref{sec:Spectral
Energy Distribution}. Note that the solid-black curves still correspond to
$L_{dbb}/L_{edd} = 0.1$, whereas the dotted-and-dashed red curves correspond to
$L_{bol}/L_{edd} = 0.1$. For details of this distinction, refer to
\dsubsecn{subsec:Diskbb}{subsec:AgnSed}.}  
\label{fig:SED}
\end{center}
\end{figure*}

To decide the value of $T(R_{in})$ for a black hole of mass $M_{BH}$, we have to consider the value of $\eta$ carefully. \citet{Zimmerman05} discuss that the $diskbb$ assumes a nonzero torque at the inner boundary of the accretion disk \citep[also see][]{Gierlinski99}, which they refer to as the `standard torque' scenario. In this scenario, the total luminosity of the disk is 
\begin{equation}
\label{eq:L_U}
	L = \frac{3G\dot{m}M_{BH}}{2R_{in}}
\end{equation}
(Equation 10 of \citet{Zimmerman05}) Thus, using Equations \ref{eq:L_eta} and \ref{eq:L_U}, we see that $\eta$ is dependant on the choice of $R_{in}$ that we make  \begin{equation}
\label{eq:Rin_eta}
	\eta = \frac{3GM_{BH}}{2c^2R_{in}}
\end{equation}
For this paper, we adopt, $R_{in} = 3R_s = 6GM_{BH}/c^2$,  for the non-spinning BH. Hence, $\eta = 0.25$.

With the above-mentioned definitions and assumptions, we have used the $diskbb$ in version 11.3 of XSPEC to generate the SED (in flux - $f_{\nu}|_{dbb}$) of the radiation from one surface of the accretion disk. We can then obtain the luminosity from one surface of the disk, by multiplying with a factor of $\pi
D^2$ and integrating over $\nu$. We have verified using $\theta = 0$, in Equation \ref{eq:Adbb}, that the ratio of 2 times this luminosity, to the Eddington luminosity is $\simeq \dot{m}/\dot{m}_{Edd}$ used in Equation \ref{eq:Temperature}. The factor 2 is to account for emission from both sides of the disk. 

After understanding this method of generating models for accretion disk radiation, we proceed to produce SEDs for the accretion disk, for a wide range of black hole masses - $M_{BH}=10^{6}-10^{12}M_{\odot}$ - at steps of 0.25 dex. For demonstration, in the left panel of Figure \ref{fig:SED}, we plot the SEDs for the different masses (as labeled), but all of them for ${L_{dbb}}/{L_{edd}}=0.1$ (solid-black lines), where $L_{dbb}$ is the luminosity obtained by integrating the accretion disc SED (from $diskbb$ in Xspec) through the entire energy range of {1 $\mu$ m - 200keV}, and then converting from flux to luminosity through a factor of $\pi D^2$. The drop in ionizing photons between the H ionization edge (912 \AA) and the He ionization edge (228 \AA) is large toward high masses. Note that the peak of the SEDs shifts towards the lower energies (or larger wavelengths) for heavier black holes. Even the SDSS filters (marked at the bottom of Figure \ref{fig:SED}, left panel) will probe a different part of the accretion disc for $M_{BH} \geq
10^{9}M_{\odot}$, than that for standard mass black holes. Even for BHs with $M_{BH}=10^{8}M_{\odot}$ at redshift z $\geq$ 3, these SDSS filters slide away the from the $f_{\nu}\sim\nu^{1/3}$ region (which for each SED is highlighted by the thicker gray lines). 

For most of the paper, we avoid the additional complication of the spin of the black hole. However, a spinning BH will have a harder SED shape in the photoionizing energy bands than a non-spinning BH of the same mass. Hence, after deriving the results for non-spinning SMBH, we perform an initial comparison of Schwarzschild and spinning BHs in Section \ref{sec:rotatingBH}. 


\subsection{The non-thermal radiation}
\label{subsec:powerlaw}

In the energy range of 2keV-10keV, AGNs have a non-thermal spectral component that is ascribed to inverse Comptonization of some of the disc photons by the hot coronal plasma surrounding the black hole \citep{Czerny87, Lightman87, Coppi92, Haardt93, Coppi99, Beloborodov99}. This spectral component can be modeled using a power-law $f(\nu) \sim
\nu^{-\alpha}$  where $\alpha$  is the spectral index \citep{Arnaud96,Zdziarski96,Zycki99}.

The power-law becomes a significant component in photoionizing the BEL ions, especially for HMBH SEDs, because the HMBH $diskbb$ moves to lower energies.

Empirically $\alpha$ depends on $L_{bol}$ \citep{Lu99,Wang04,Shemmer06}, though there is not complete consensus \citep{Trakhtenbrot17}. \citet{Brightman13} gives a phenomenological relation between the spectral index and $L_{bol}$ as,\begin{equation}
\label{eq:Brightman}
\Gamma = \alpha +1 = (0.32 \pm 0.05) \times log (\frac{L_{bol}}{L_{Edd}}) + (2.27 \pm 0.06)
\end{equation}where $L_{bol}$ is the total `observed' Luminosity between 1$\mu$ and hard X-rays (200 keV). Note that the entire 1$\mu$ to 200 keV range cannot be observed - sometimes because of the natural extinction (by Galactic dust and neutral hydrogen) in the $\sim 10 - 200$ eV; and sometimes because of bad data in other wavelengths, for the samples that \citet{Brightman13} considered. Hence they had made judicious extrapolations and modeling, where required.  We then define $L_{bol}$ as the sum of the luminosities from both the SED components - the accretion disc and the power-law, in the same energy range. Using Equation \ref{eq:Brightman}, we see that the spectral indices for power-law components will be  0.78, 0.95, 1.10 \& 1.27 respectively, for the different $L_{bol}/L_{edd}=$0.03, 0.1, 0.3 and 1.0 ratios that we plan to investigate in this paper. Note that, in this paper, we shall use ${L_{bol}}/{L_{edd}}$ as proxy for the accretion rate. Using these spectral energy indices we construct power-law component and introduce exponential cut-offs at both the ends.\begin{equation}
\label{eq:power-law}
f(\nu) \sim \nu^{-\alpha} \times e^{\frac{-10eV}{\nu}} \times e^{-\frac{\nu}{200keV}} 
\end{equation}

Once the power-law component is constructed, we need to devise a scheme for its relative normalisation with respect to the $diskbb$ component.

\begin{table}
\caption{Maximum $M_{BH}$ for which 2500 \AA $ $ falls on the linear part of their SED.}
\label{tab:MbhRange}
\centering
\begin{tabular}{l l}
\hline
\hline
$\frac{L_{bol}}{L_{Edd}}$ & log($\frac{M_{BH}}{M_{\odot}}$) \\
\hline
0.03 & 7.75\\
0.1 & 8.25\\
0.3 & 9.00\\
1.0 & 9.25\\
\hline
\hline
\end{tabular}
\end{table}

\subsection{The multicomponent broad-band AGN SED}
\labsubsecn{subsec:AgnSed}

Conventionally, the nominal UV to X-ray slope, $\alpha_{OX}=-0.384$ log $(L_{2keV}/L_{2500 \textup{\scriptsize{\AA}}})$ \citep{Tananbaum79} is used to relate the luminosities at 2500\AA $ $ and 2 keV, where the accretion disc radiation predominantly determines the luminosity at 2500\AA, while the X-rays at 2 keV mostly comes from the power-law component. For the BHs in the `standard AGN mass range',  $M_{BH} \lesssim 10^{8} M_{\odot}$, and $L_{bol}/L_{edd} \gtrsim 0.1$, 2500\AA $ $ falls on the part of the $diskbb$ SED, where $f_{\nu} \sim \nu^{1/3}$ (thick highlighted gray lines in the left panel of Figure \ref{fig:SED}). However, for HMBHs the peak of the $diskbb$ emission moves to lower energies and $f_{\nu} \sim \nu^{1/3}$ may not be satisfied at 2500\AA, as demonstrated in left panel of Figure \ref{fig:SED}. The highest possible black hole mass, for which 2500\AA $ $ deviates from $f_{\nu} \sim
\nu^{1/3}$, is also a function of the accretion rate, $L_{bol}/L_{edd}$. For the different $L_{bol}/L_{edd}$ ratios, used in this paper, in Table \ref{tab:MbhRange}, we have listed the values of the highest mass for which 2500\AA $ $, in the corresponding SEDs, falls on the $f_{\nu} \sim \nu^{1/3}$ part of the $diskbb$ SED. Thus, for HMBHs, 2500\AA $ $ represents a physically different part of the accretion disc than in case of the standard mass black holes. The reason we pay due attention to the nature of the SED at 2500 \AA $ $, is because this wavelength has been used as a reference wavelength in many multiwavelength AGN SEDs, even for the more recent, robust, SED investigations, where Ultraviolet and optical luminosities are related to each other. \citet{Lusso16} examined a large sample of 159 AGN and found a tight relationship between the luminosities at 2500\AA $ $ and 2 keV. \begin{equation}
\label{eq:Lusso&Risaliti}
log(L_{2keV}) = 0.638 \times log(L_{2500 \AA}) + 7.074
\end{equation}While they use the same wavelengths, as used by the definition of $\alpha_{OX}$, this form of the relationship directly relates the observables, namely the fluxes.

\begin{table}
\caption{Constants of the $log(\frac{M_{BH}}{M_\odot})-\frac{L_{pl}}{L_{dbb}}$ best-fit equation.}
\label{tab:constants}
\centering
\begin{tabular}{l l l l}
\hline
\hline
$\frac{L_{bol}}{L_{Edd}}$ & $a_{1}$ & $a_{2}$ & $b_{2}$ \\
\hline
0.03 & -8.21633  & 165.093  & -3.58592 \\
0.1 & -1.35708   &21.282 & -0.523004\\
0.3 & -1.30806 & 23.599 & -2.24452  \\
1.0 & -0.674781  & 12.5449  & -2.31223 \\
\hline
\hline
\end{tabular}
\end{table}

\begin{figure}
\begin{center}
\includegraphics[width=0.45\textwidth, trim= 0 10 0 0]{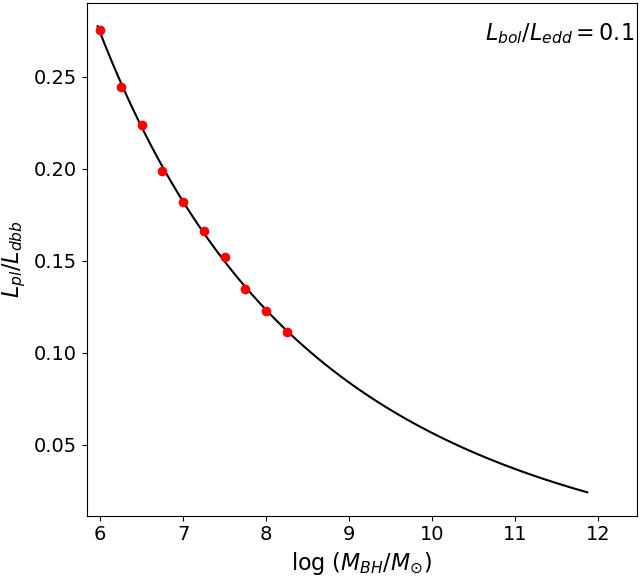}
	\caption{$L_{pl}/L_{dbb}$ ratios as a function of mass ratios for
	$L_{bol}/L_{edd}=0.1$. Red circles are the ratios obtained by using Eq.
	\ref{eq:Lusso&Risaliti} to derive $L_{2keV}$. The red circle with the
	highest mass represents the highest mass black hole for which $2500$
	\AA $ $ falls within the $f_{\nu} \sim \nu^{1/3}$ part of the SED.
	Black line represents the best-fitting curve through the red circles
	and is further extrapolated to find the $L_{pl}/L_{dbb}$ ratios for
	higher masses, using the fit function Eq. \ref{eq:extrapolation}.}
\label{fig:extrapolation}
\end{center}
\end{figure}

For each $L_{bol}/L_{edd}$, we first choose the range of $M_{BH}$ for which the 2500\AA $ $ falls on the part of the $diskbb$ SED, where $f_{\nu} \sim \nu^{1/3}$. For example, as seen from Figure \ref{fig:SED} (left panel) and Table \ref{tab:MbhRange}, for $L_{dbb}/L_{edd} = 0.1$ this mass range corresponds to $M_{BH}=10^{6}M_{\odot}$ to $10^{8.25}M_{\odot}$. For each of these SEDs, we then scale the power-law component relative to the $diskbb$ as follows. For each value of $M_{BH}$, we start with $L_{dbb}/L_{edd}$ = desired $L_{bol}/L_{edd}$ = 0.1 (say). The XSPEC generated $diskbb$ spectra give us the value of $L_{2500\AA}$. Thus, we can use that value in Equation \ref{eq:Lusso&Risaliti} to derive $L_{2keV}$, which gives us the required normalisation for the power-law  omponent. We add the luminosity $L_{dbb}$ in the $diskbb$ and the luminosity $L_{pl}$ in the normalised power-law component and check if $L_{bol}/L_{edd}$ = 0.1. If $L_{bol}/L_{edd} \neq 0.1$, then we iterate by changing $L_{dbb}/L_{edd}$ slightly, and repeating the process. The iteration continues, until, for the given $M_{BH}$, we achieve $L_{bol}/L_{edd} = (L_{dbb} + L_{pl})/L_{edd} = 0.1$. At this stage, we calculate the luminosity ($L_{pl}$) in the power-law component and the luminosity in the ($L_{dbb}$) in the $diskbb$ component, whence we can derive the ratio ${L_{pl}}/{L_{dbb}}$. In Figure \ref{fig:extrapolation}, the red circles correspond to the mass range for which the ratio ${L_{pl}}/{L_{dbb}}$ was derived using the aforementioned method. The solid black line joining the red circles is then extrapolated using a smooth, power series extrapolation,\begin{equation}
\label{eq:extrapolation}
\frac{L_{pl}}{L_{dbb}} = \frac{a_{1}}{\frac{M_{BH}}{M_{\odot}}} + \frac{a_{2}}{(\frac{M_{BH}}{M_{\odot}}-b_{2})^{2}} 
\end{equation}where the values of constants are provided in the Table \ref{tab:constants}. Thus, for the higher mass black holes, as well, we get the scaling of the power-law component relative to the $diskbb$ component, which can then be used to construct the corresponding SEDs. We have thus, built a suite of SEDs (100 of them) for $L_{bol}/L_{edd} = 0.03, 0.1, 0.3, 1.0$ with the range of mass varying from $M_{BH}=10^{6}M_{\odot}$ to $10^{12}M_{\odot}$ in intervals of 0.25 dex. The right panel of the Figure \ref{fig:SED} shows the optical to X-rays SEDs (red dotted-and-dashed lines) obtained thus, for $L_{bol}/L_{edd} = 0.1$, where the steeply falling (relative to the power-law) higher energy tails of the $diskbb$ are shown in solid black lines. Above $M_{BH} \sim 10^{9} M_{\odot}$ the power-law dominates the main ionizing continuum relevant for the BEL ions. For lower accretion rate ($L_{bol}/L_{edd}$ <0.1) this statement is true for even lower black hole mass.

One might argue that Equation \ref{eq:Lusso&Risaliti} does not distinguish between super and hyper massive black holes and so this equation should be used directly, even for the HMBHs and we should not extrapolate. However, these extrapolations, which become important only for the HMBHs, are based on the physics of accretion discs. Further, in \secn{sec:Lusso}, we show that if we had directly used Equation \ref{eq:Lusso&Risaliti} for the highest masses, the differences in the results would be minor, and effectively has no difference in the qualitative inferences.


\section{Photoionization Calculation}
\label{sec:CLOUDY Calculation}

For each of 100 SEDs (25 masses $\times$ 4 $L_{bol}/L_{edd}$ values)), we use the C17.00 version of CLOUDY \footnote{{http://www.nublado.org/}} \citep{Ferland17}, to calculate the EWs of many emission lines for a wide range of the parameters  $\Phi(H)$, $n_{H}$  and $N_{H}$. $\Phi(H)$ is the ionizing incident photon flux $\Phi(H) = Q(H)/4 \pi r^{2}$, where Q(H) is the number of Hydrogen ionizing photons.  

Though CLOUDY gives EW of each line as a function of $\Phi(H)$, $n_{H}$, the parameter $\Phi(H)$ can easily be interchanged by $r$. Hence, we can express (following Hazy 2) that the flux of the emission line from a cloud of density $n_{H}$ and at a distance $r$ is 
\begin{equation}
\label{eq:EWtoF}
F(r,n_{H})\sim EW(r,n_{H}) \frac{\nu F_{\nu}^{c}}{\lambda}
\end{equation} 
where $\lambda$ is the central wavelength of the line and $\nu F_{\nu}^{c}$ is the incident continuum at $\lambda$. $F(r,n_{H})$ can then be used in Eq. \ref{eq:LOCLine3} to calculate the total line luminosity $L_{line}$, as described in the next section.

For the calculation of BELs, the parameters are stepped over a grid of values as shown in Table \ref{tab:BELparameter}, a total of 845 CLOUDY runs (for each SED). These ranges are based on \citet{Baldwin95}, \citet{Korista97a} and the range of $N_{H}$ on \citet{Chakravorty14}. We have calculated the EWs of all the 42 prominent quasar emission lines listed in \citep{Korista97a}, out of which only 5 strongest lines are demonstrated here. 

\begin{table}
\caption{BEL parameter ranges}
\label{tab:BELparameter}
\centering
\begin{tabular}{l l l l}
\hline
\hline
 & min & max & $\Delta$ \\
\hline
log $\Phi (H)$ & 18 & 24 & 0.5\\
log $n_{H}$ & 8 & 14 & 0.5\\
log $N_{H}$ & 21 & 23 & 0.5\\
\hline
\hline
\end{tabular}
\end{table}

For the calculation of NELs, the ranges in Table \ref{tab:NELparameter} were used. We adopted these ranges based on \citet{Ferguson97}. We have calculated the EWs of all the 23 prominent narrow emission lines listed in \citet{Ferguson97}, out of which only 6 strongest lines are demonstrated here. For NELR, the presence of dust can play some role in determining the gas composition and hence the strength of lines. \citet{Ferguson97} demonstrate the effect of the presence of dust in detail. However, they also find a combination of parameters, where the dust-free gas produces very similar line strengths (matching observations) as a dusty gas. For the sake of simplicity, we adopt those parameters and work with dust-free gas for the NELR. The appropriate parameters are discussed in more details in Section \ref{sec:Narrow Emission Lines}.

\begin{table}
\caption{NEL parameter ranges}
\label{tab:NELparameter}
\centering
\begin{tabular}{l l l l}
\hline
\hline
 & min & max & $\Delta$ \\
\hline
log $\Phi (H)$ & 10 & 20 & 0.5\\
log $n_{H}$ & 2 & 10 & 0.5\\
log $N_{H}$ & 21 & 23 & 0.5\\
\hline
\hline
\end{tabular}
\end{table}

Thus for the total number of SEDs explored in this paper, along with the calculations for each of the physical parameters of the gas properties $\Phi(H)$, $n_{H}$  and $N_{H}$, we ran a total of 256,240 CLOUDY models. For SEDS with low accretion rate, we have done CLOUDY calculations only up to $M_{BH}=10^{11.5}M_{\odot}$ due to numerical convenience. This huge theoretical data set was then used to calculate
the optimal line strengths as discussed in the next Section (\ref{sec:LOC
Model}).

\begin{figure*}
\begin{center}
\includegraphics[width=0.9\textwidth, trim = 0 15 0 0]{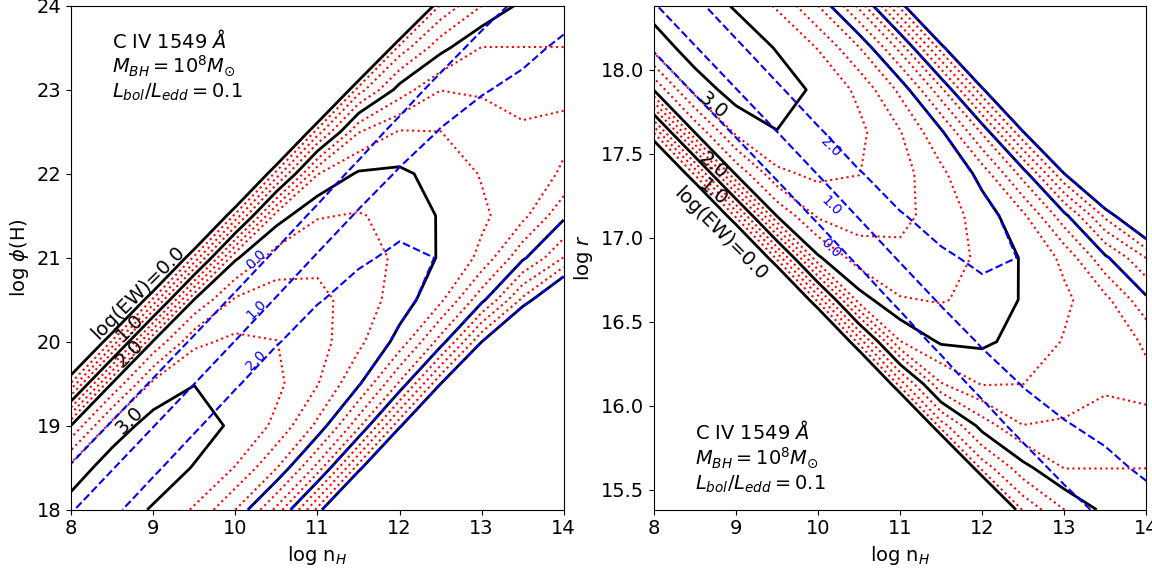}
\caption{Contour plot for EW of C IV (1549 \AA)  in the $n_{H}-\Phi(H)$ plane
	(left panel) and $n_{H}-R_{cloud}$ plane (right panel). The solid black
	lines are contours at steps of 1 dex and dotted red lines are at the
	steps of 0.25 dex, for log$N_{H}$=23. The dashed blue lines are
	contours at steps of 1 dex, for log$N_{H}$=21.  Originally, Baldwin et
	al. (1995) and Korista et al. (1997) presented the efficiency of line
	emission for different BELs as a function of $n_{H}$ and $\Phi$ (as in
	the left panel). Ferguson et al. (1997) made a similar study on NELs;
	efficiency was presented as a function of $n_{H}$ and $R_{cloud}$ (as
	in the right panel) instead. For the sake of consistency, between the
	BEL and NEL analysis we stick to $R_{cloud}$ instead of $\Phi$,
	throughout the rest of this paper. Here, we represent both planes, to
	demonstrate that its a mere inversion of the contours, when the
	y-parameter of the plane is changed from $\Phi$ to $R$. }  
\label{fig:BLRContour}
\end{center}
\end{figure*}

\section{Locally Optimally Emitting Clouds}
\label{sec:LOC Model}

Strong BELs are the identifying feature of AGN and study of them gives information on their central engine. The BELs are also used to estimate the chemical composition of the hosting galaxy and their EWs are even used in estimating the black hole mass. 

Photoionization models of the broad line region are extremely relevant to the study of AGN \citep{Netzer92}. Early BLR models were quite limited and used to assume the lines were emitted by the single gas cloud, one with a single ionization parameter, column density and gas density \citep{Kwan81}. These models became untenable when reverberation mapping observations showed the presence of a wide distribution of emission-line cloud properties with distance from the ionizing continuum. This development prompted a multi-cloud model with a wide range of spatial and density distributions \citep{Baldwin95}.

The ``locally-optimally emitting cloud" (LOC) model was introduced by \citet{Baldwin95}. This model established that any given BEL is most intensely emitted by gas with a particularly narrow range of ionization parameter and density and thus a weighted average over spatial and density distribution includes the gas with optimal parameters for each of the observed lines. \citet{Ferguson97} further extended this model to predict the line strengths of Narrow Emission Lines (NELs). 

The total line luminosity emitted by the entire set of clouds at various radial distances and densities is given by,\begin{equation}
\label{eq:LOCLine1}
L_{line} \propto \int \int r^{2} F(r,n_{H}) f(r)g(n_{H}) dn_{H} dr 
\end{equation}where $f(r)$ and $g(n_{H})$ are the spatial and density distribution functions respectively. For simplicity we assume them to be power laws as in \citet{Ferguson97}. We have further normalised the $L_{line}$ over the entire spatial and density distribution functions .\begin{equation}
\label{eq:LOCLine3}
L_{line} = 0.24 \, \frac{\int \int r^{2} F(r,n_{H}) r^{\gamma} n_{H}^{\beta} dn_{H} dr }{\int \int {r^{\gamma}} n_{H}^{\beta} dn_{H} dr }
\end{equation}

Similarly, the weighted average of the EWs ($\mathcal{EW}$) of each line emitted by the entire set of clouds at various radial distances and densities is calculated using,\begin{equation}
\label{eq:LOCEW1}
\mathcal{E} \mathcal{W} = 0.24 \, \frac {\int \int EW (r, n_{H}) {r^{\gamma}} n_{H}^{\beta} dn_{H} dr }{\int \int {r^{\gamma}} n_{H}^{\beta} dn_{H} dr }
\end{equation}
 
The successes and the limitations of LOC model are discussed in detail by \citet{Leighly07}. Limitations of LOC model include that (i) some parameters, like the spatial and density indices cannot be physically interpreted and (ii) a number of physical effects are not included in the model, a significant one being self-shielding. However, for the purpose of this paper, the LOC model is sufficient.

The factor 0.24 in Equations \ref{eq:LOCLine3} and \ref{eq:LOCEW1} represents a nominal constant covering factor for the broad line emitting clouds \citep[following][see Figure 1]{Leighly07}. Covering fraction is a very complex issue, varying from source to source. Hence its theoretical generalisation over a wide range of mass and accretion rates of BHs is difficult and beyond the scope of this paper. Hence we maintain this value of 0.24 throughout the paper, even for narrow line emitting clouds. Please note that any different constant covering factor would simply scale the $\log(\mathcal{EW})$s in the vertical direction, in our subsequent Figures, altering none of the trends demonstrated. Further, a constant covering factor will absolutely have no effect of the line ratios, presented later.

\section{Broad Emission Lines}
\label{sec:Broad Emission Lines}

\begin{figure}
\begin{center}
\includegraphics[width=0.45\textwidth, trim = 0 10 0 0]{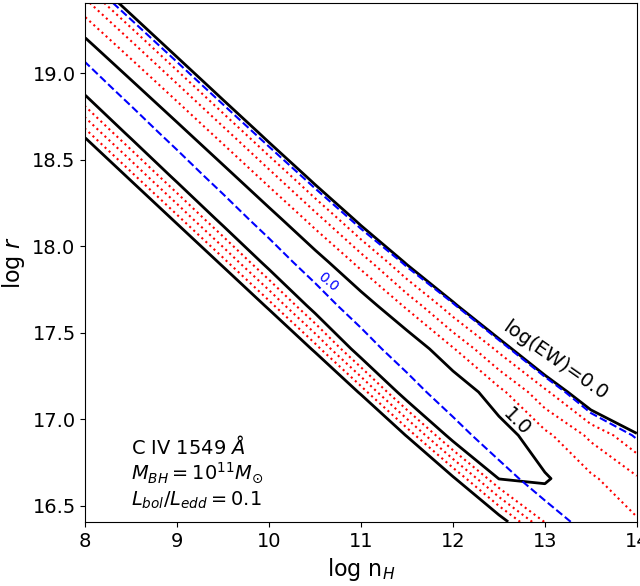}
	\caption{Contour plot for EW of C IV (1549 \AA) in the
	$n_{H}-R_{cloud}$ plane for $M_{BH}=10^{11}M_{\odot}$. The efficiency
	of the line emission has decreased greatly when compared to
	$M_{BH}=10^{8}M_{\odot}$ (Figure \ref{fig:BLRContour}). The different
	line styles and colours represent the same variation of physical
	parameters as in Figure \ref{fig:BLRContour}.}  
\label{fig:BLR11}
\end{center}
\end{figure}

From the CLOUDY photoionization calculations, we derive EWs for all the
prominent BELs. The EWs are measured relative to the incident continuum at 1215
\AA $ $ which is then normalised to the central wavelength of each line. In
Figure \ref{fig:BLRContour} we represent the iso-contours of calculated EWs (in
log) for C IV 1549 \AA, one of the strongest known broad emission lines on the
log$n_{H}$-log$\Phi(H)$ (left panel) and log$n_{H}$-log$r$ (right panel) planes
for $M_{BH}=10^{8.0} M_{\odot}$ and $L_{bol}/L_{Edd}=0.1$. We see that our
(solid-black and dotted-red) contour plots match closely with the contours
presented in \citet{Baldwin95} and \citet{Korista97a}, for
$logN_{H}=23$. The slight quantitative differences are expected because the
earlier authors used a slightly different SED compared to the one that is used
to get Figure \ref{fig:BLRContour}. The diagonal lines in the contours with a
slope of $45^{\circ}$ are of constant ionization parameter U =
$\Phi(H)$/$cn_{H}$, which is a measure of the recombination rate at the face of
the cloud. C IV being a collisionally excited line shows a band of constant U
lines where there is efficient emission.  At high $n_{H}$ and $\Phi(H)$,
contours start to turn over indicating thermal heating of the gas instead of
photoionization. To understand the effects of column density we point to the
isocontours for $logN_{H}=21$ in dashed blue line, in Figure
\ref{fig:BLRContour}. When compared to the solid-black $logN_{H}=23$ contours
(at same 1 dex separation), the ones with lower $N_H$ show that for highly
ionized gas, we need high column density gas to yield emission lines of any
reasonable strength. In Figure \ref{fig:BLR11}, we represent the contour for
$M_{BH}=10^{11}M_{\odot}$ to compare the efficiency of line emission at higher
masses. While the highest EW achieved for the $M_{BH}=10^{8}M_{\odot}$ SED is
$10^{2.5}$, that for the $M_{BH}=10^{11}M_{\odot}$ SED is $10^{1.25}$. Thus,
everything else remaining the same, the line becomes fainter by 1.25 dex. 

\begin{figure}
\begin{center}
\includegraphics[width=0.47\textwidth, trim = 0 15 0 0]{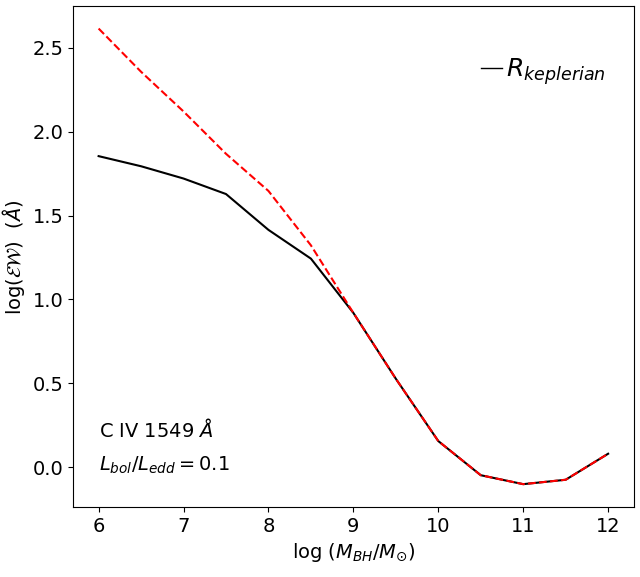}
\caption{$\mathcal{EW}$ of C IV (1549 \AA) as a function of $M_{BH}/M_{\odot}$.
	Dashed line represents $\mathcal{EW}$ calculated using LOC integration
	(Equation \ref{eq:LOCEW1}). The black solid line represents
	$\mathcal{EW}$ after limiting the outermost radius of cloud with
	$v_{min}=1000km s^{-1}$.}  
\label{fig:EWC4}
\end{center}
\end{figure}

To assess the contribution from all the clouds spread across a range of
distance and having a range of density, we apply the LOC model mentioned in
Section \ref{sec:LOC Model} for our calculations. We use Eq. \ref{eq:LOCLine3}
and Eq. \ref{eq:LOCEW1} over the entire range of $r$ and $n_{H}$. \citet{Baldwin97}
 suggested that the $L_{line}$ is only weakly sensitive to the radial and
column density distributions as long as $\gamma$ in Eq. \ref{eq:LOCLine3} is $>
-1$. \citet{Baldwin97} further shows that $\beta=-1$ is strongly suggested by
observations which is further confirmed by \citet{Hamann02}. Therefore, we
use a constant power index of -1 for both, $n_{H}$ and $r$ in our calculations
to get $L_{line}$ and $\mathcal{EW}$. Figure \ref{fig:EWC4} shows
the resultant log($\mathcal{EW}$) as a function of $M_{BH}$ for the
BEL C IV 1549 \AA, for $L_{bol}/L_{Edd}=0.1$ (red dotted line).

At $10^{4}$K, thermal widths are only about 10 km s$^{-1}$. The observed widths
in BELs ($\gtrsim$ 10 km s$^{-1}$) should, therefore, be due to organised high
velocity flows of clouds with Keplerian, or at least, virialised velocities.
This allows us to define maximum radius for BLR as, \begin{equation}
\label{eq:keplarian}
 R_{Keplerian}= G M _{BH}/ v_{min}^{2}
\end{equation}where $v_{min}$ is the lowest observed velocity of the cloud. Introducing
$R_{Keplerian}$ cut-off within CLOUDY calculated line strengths, results in a
significant change in $\mathcal{EW}$ at the lower mass end as in
\citet{Chakravorty14}. In Figure \ref{fig:EWC4}, the black solid line
corresponds to $\mathcal{EW}$ after imposing a lower limit at $v_{min} \geq$
1000 km s$^{-1}$. For the rest of the paper, we will always calculate
$\mathcal{EW}$ and $L_{line}$ for (all) BELs, maintaining this lower limit of
$v_{min}$.

\begin{figure*}
\begin{center}
\includegraphics[width=\textwidth, trim = 0 20 0 0]{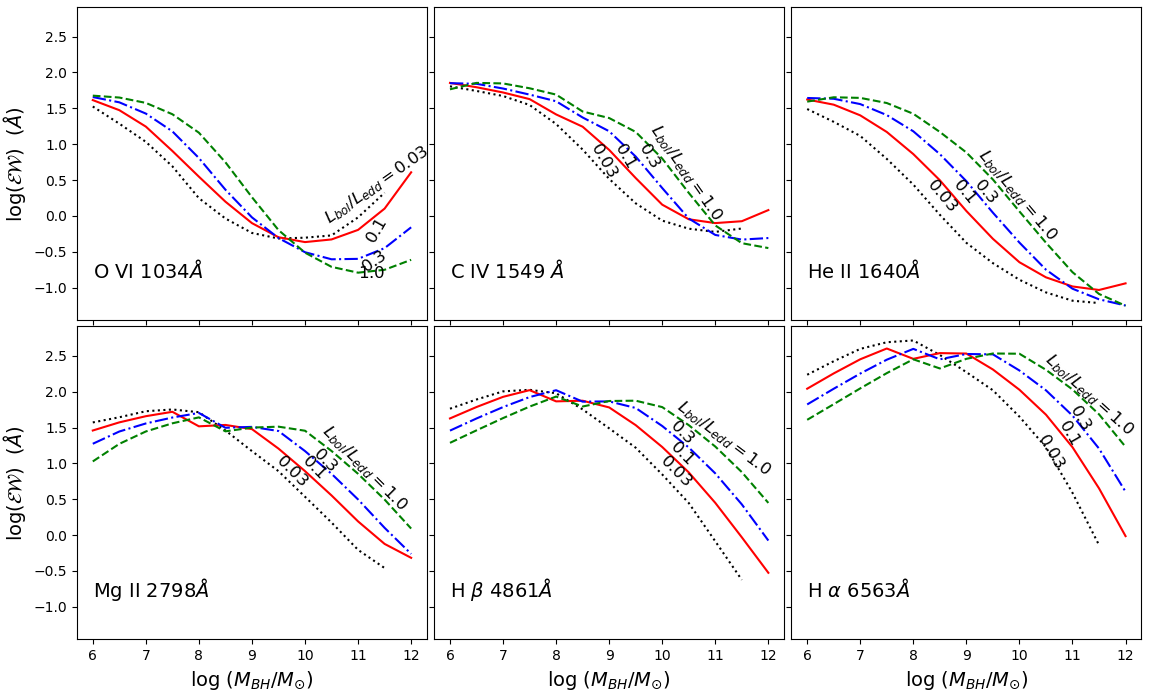}
	\caption{Evolution of $\mathcal{EW}$ of few prominent BELs over range of masses and accretion rates.}
\label{fig:BEL_EW}
\end{center}
\end{figure*}

In the different panels of Figure \ref{fig:BEL_EW}, we extend the calculation
and show how the line strength $\mathcal{EW}$ of six of the
prominent BELs such as O VI 1034 \AA, C IV 1549 \AA, He II 1640 \AA, Mg II 2798
\AA, H$\beta$ 4861 \AA $ $ and H$\alpha$ 6563 \AA $ $ vary when accretion rate is
varied over the range $0.03 \leq L_{bol}/L_{edd} \leq 1.0$.  

For C IV 1549 \AA $ $ (top middle panel of Figure \ref{fig:BEL_EW}) at
$L_{bol}/L_{edd}=0.1$ the  $\mathcal{EW} = 26$ for the $10^8 M_{\odot}$ BH,
which is in excellent agreement with what Figure 1 of \citet{Leighly07}
predicts for the same emission line, for the nominal model (with $C_f = 0.24$).
The  $\mathcal{EW}$ of this line remains quite constant for $10^{6}M_{\odot} <
{BH} < 10^{8}M_{\odot}$ and then drops quickly to 1\% of the peak by $\sim
10^{10} M_{\odot}$. O VI 1034 \AA, He II 1640 \AA and Mg II 2798 \AA $ $ all
show similar $\sim$2 dex decrease in $\mathcal{EW}$ from $10^{8}M_{\odot}$ to
$10^{10}M_{\odot}$.  H$\beta$ 4861 \AA $ $ and H$\alpha$ 6563 \AA $ $ show similar
behavior from $10^{9}M_{\odot}$ to $10^{11}M_{\odot}$. 

In Figure \ref{fig:BEL_EW} notice the distinction between the $\mathcal{EW}$
profiles of the high ionization O VI, C IV, and He II (top panels) to those of
the low ionization Mg II, H$\beta$ and H$\alpha$ lines (bottom panels). The low ionization lines maintain quite constant
$\mathcal{EW}$ up to $\sim 10^{10} M_{\odot}$ while the high ionisation lines
begin to drop in $\mathcal{EW}$ by $\sim 10^8 M_{\odot}$, for the BHs with
accretion rate $L_{bol}/L_{edd}=0.1$.  The aforementioned mass ranges mentioned
for the $\mathcal{EW}$ profiles vary when the accretion rate changes, but the
qualitative behaviour of the different lines remains the same. Also, the high
ionization profiles decrease with increase of mass, up to a critical value of
BH mass, hit a minimum and then invert the trend and start increasing with
mass. On the other hand, the low ionization lines profiles show increase (or
almost constant values) of $\mathcal{EW}$ with mass increase, hit a maximum and
then drops with increase in mass. Such differing evolution of the
$\mathcal{EW}$s will also have consequences on how the luminosities of these
lines would evolve as a function of the BH mass, particularly at the high mass
end. 

\begin{figure*}
\begin{center}
\includegraphics[width=0.9\textwidth, trim = 0 12 0 10]{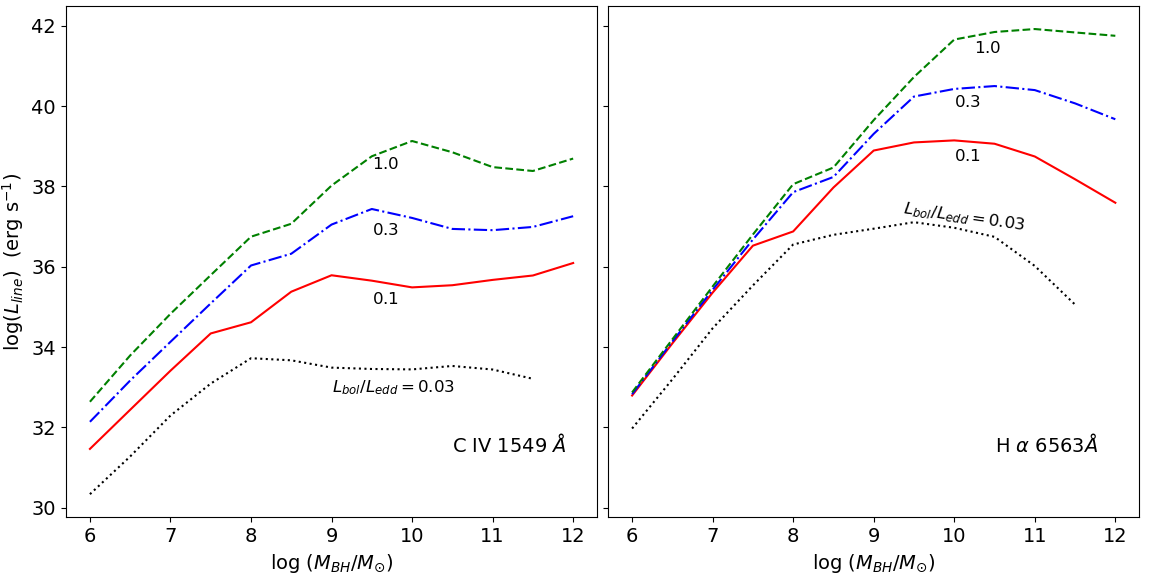}
\caption{$L_{line}$ of C IV 1549 \AA $ $ and H$\alpha$  6563 \AA $ $ as a function of $M_{BH}/M_{\odot}$ calculated using Eq. \ref{eq:LOCEW1}.}  
\label{fig:EWC4lum}
\end{center}
\end{figure*}

We take C IV 1549 \AA $ $ as a representative of the high ionization lines and
H$\alpha$ 6563 \AA $ $ as a representative of the low ionization lines and plot
their luminosities as a function of mass in Figure \ref{fig:EWC4lum}. At the
high mass end, while the luminosity of the C IV line remains constant, the
luminosity of H$\alpha$ line drops, sometimes by even an order of magnitude.
\textit{Thus the low ionization lines are not the best tracers to look for
while searching for the HMBH using the current optical telescopes. This result
has implications for the most commonly done BH searches that rely on optical
surveys like SDSS, which in their turn, rely heavily on the H$\beta$ 4861 \AA $
$ and H$\alpha$ 6563 \AA $ $ lines. However, the results presented here may
serve as a benchmark for emission line studies, using the next generation 30 m
class optical telescopes.} 

\begin{figure}
\begin{center}
\includegraphics[width=0.44\textwidth, trim= 0 12 0 10]{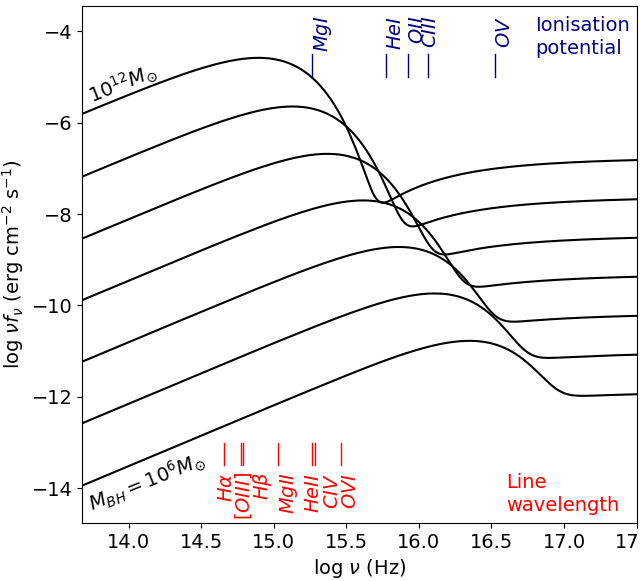}
	\caption{The line wavelength of the 5 strongest BELs (marked at the
	bottom) and the ionization potentials of the lower ions (marked on the
	top) are compared against the shape of the broad band SEDs of different
	BHs, for $L_{bol}/L_{edd} = 0.1$. } 
\label{fig:ionization}
\end{center}
\end{figure}

The distinction between the behaviour of the high and the low ionization lines
is readily understood from Figure \ref{fig:ionization}.  The ionization
potentials (IP) of O V and C III, which would result in the formation of O VI
and C IV ions, are the highest, among the BEL ions/elements discussed. Note
that the IP of these ions are at an interesting energy range - one where the
accretion disc is more dominant for the SEDs of the lower mass ($M_{BH}
\lesssim 10^8 M_{\odot}$) black holes, but the power-law radiation component is
more dominant for the higher mass BHs. For each ion (high ionization lines),
the narrow mass range, where the transition of the relative dominance happens,
is where the $\mathcal{EW}$ turns over. For ions (low ionization lines), where
there is no significant switch of the dominating SED component, the
$\mathcal{EW}$ simply declines with an increase in BH mass. In fact, for the
H$\alpha$ and H$\beta$ lines from {\bf neutral}, but excited Hydrogen, of
course IP is not the deciding factor. Hence, they follow the simple rationale
that the $\mathcal{EW}$ increases with the mass (and hence the luminosity) of
the BH until their line energies go above the $f(\nu) \sim \nu^{1/3}$ part of
the BH SED - which happens because the accretion disc peak and hence the
$\nu^{1/3}$ (part of the SED) recedes to lower energies with an increase in BH
mass. Thus, at the corresponding BH mass, the H$\alpha$ and H$\beta$
$\mathcal{EW}$s start dropping. This discussion thus illustrates the fine
sensitivity that line strengths have on the shape of the illuminating SED.
Further, this sensitivity becomes evident in our study because we are probing a
mass range of BHs where the SEDs transition through the energy range of the
relevant IPs and line energies - a very fortunate natural coincidence indeed!    

Figure \ref{fig:BEL_EW} further shows that the turn around of log($\mathcal{E}
\mathcal{W}$) in O VI 1034 \AA, C IV 1549 \AA $ $ for higher masses makes it
impossible to determine the mass of the HMBH accurately using just
log($\mathcal{EW}$) of these strong BELs. For the other emission lines, similar
degeneracy exists at the lower mass end. Hence, to remove this degeneracy we
have calculated line ratios which we discuss in Section \ref{subsec:BelLR}.

\begin{figure}
\begin{center}
\includegraphics[width=0.45\textwidth, trim= 0 10 0 10]{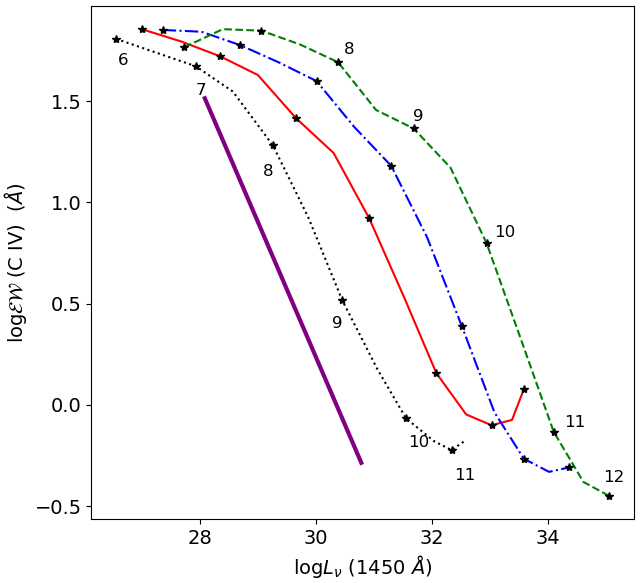}
	\caption{Demonstration of the Baldwin effect using the CIV 1549 \AA $ $
	emission line. The left-most thick solid magenta line is the predicted
	Baldwin effect relationship $\mathcal{EW} \sim L^{-2/3}_{\nu}(1450
	\AA)$, with arbitrary normalisation. The rest of the lines are the
	$\mathcal(EW)$ profiles of the C IV line, having the same line-colour
	description as in Figure \ref{fig:BEL_EW} - from accretion rate of 0.03
	to 1.0, from left to right. The numbers on each profile indicate the
	mass of the BH, in $log(M_{BH}/M_{\odot})$.} 
\labfig{fig:BaldwinEffect}
\end{center}
\end{figure}

We are in a position to investigate if the Baldwin Effect \citep{Baldwin77,
Zheng93, Shemmer15} holds true for the HMBHs and gain some insight on the mass
range or range of accretion rates for which this effect remains valid. Hence,
we plot the $\mathcal{EW}$ of CIV 1549 \AA $ $ as a function of $L_{\nu}(1450
\AA)$ in \fig{fig:BaldwinEffect}. The left-most thick solid magenta line in
\fig{fig:BaldwinEffect} shows the relationship, $\mathcal{EW} \sim
L^{-2/3}_{\nu}(1450 \AA)$, predicted by \citet{Baldwin77} between the C IV line
strength and the Luminosity at $1450 \AA$. Clearly, the low mass BHs ($M_{BH} <
10^7 M_{\odot}$) do not obey this relationship. 
 However, for higher masses
($10^8 \leq M_{\odot} \leq 10^{11}$), the match with the slope of -2/3 (= -
0.66) is good and improves for higher accretion rates; albeit the mass range
where the match is better, also shifts to the higher values with the increase
in accretion rate. For $L_{bol}/L_{Edd} = 0.03$, the slope is -0.51 for mass
range $10^7 - 10^{10} M_{\odot}$, for 0.1, the slope is -0.53 for $10^8 - 10^{11}$, for
0.3, the slope is -0.66 for $10^9 - 10^{11}$ and finally, for 1.0, the slope is
-0.63 for $10^9 - 10^{12}$.
Beyond $10^{11} M_{\odot}$, the Baldwin effect
disappears as the $\mathcal{EW}$ of the C IV line starts to strengthen again,
especially, at low accretion rates. Thus, we show that the Baldwin effect is a
natural result of the changing accretion disk SED with BH mass over 2 - 4
orders of magnitude.


Note that the discussion of the behaviour of emission line strengths in this
section is for non-spinning black holes. To understand how BH spin may play a
role (particularly for the OVI and the CIV lines) in this discussion, refer to
Section \ref{sec:rotatingBH}. 


\section{Narrow Emission Lines}
\label{sec:Narrow Emission Lines}

\begin{figure*}
\begin{center}
\includegraphics[width=0.9\textwidth, trim = 0 12 0 0]{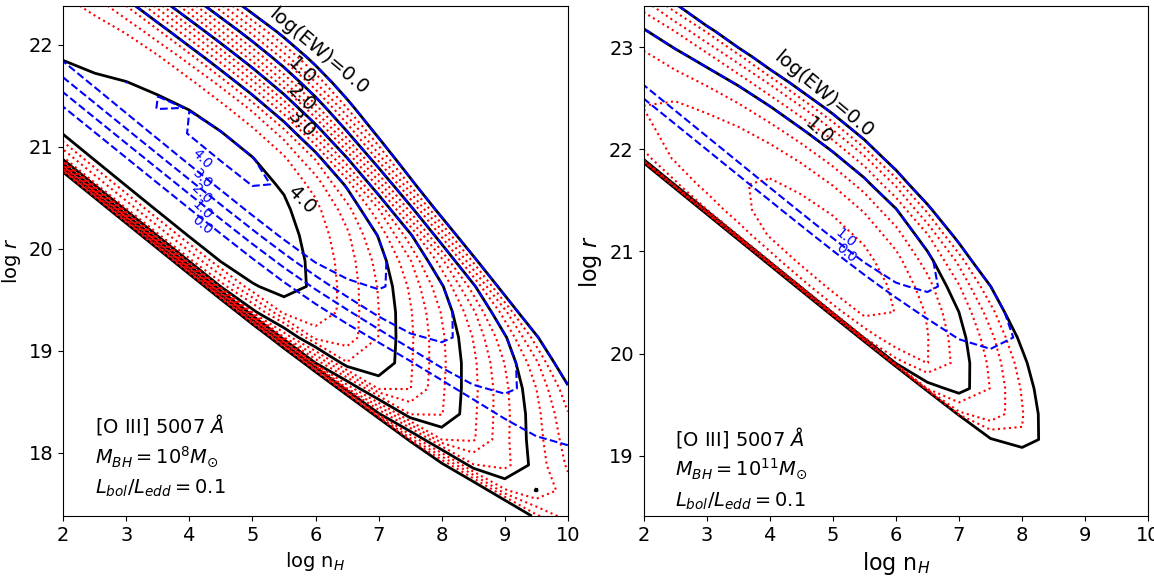}
\caption{Contour plot for EW of [O III] 5007 \AA $ $ in the $n_{H}-r$ plane for $M_{BH}=10^{8}M_{\odot}$ (left panel) and $10^{11}M_{\odot}$ (right panel). The solid black lines are contours at steps of 1 dex and dotted red lines are at the steps of 0.25 dex, for log$N_{H}$=23. The dashed blue lines are contours at steps of 1 dex, for log$N_{H}$=21. It is evident that the efficiency of the line emission has decreased greatly in $M_{BH}=10^{11}M_{\odot}$ when compared to $M_{BH}=10^{8}M_{\odot}$.}
\label{fig:NLRContourR}
\end{center}
\end{figure*}

\begin{figure}
\begin{center}
\includegraphics[width=0.45\textwidth, trim = 0 12 0 0]{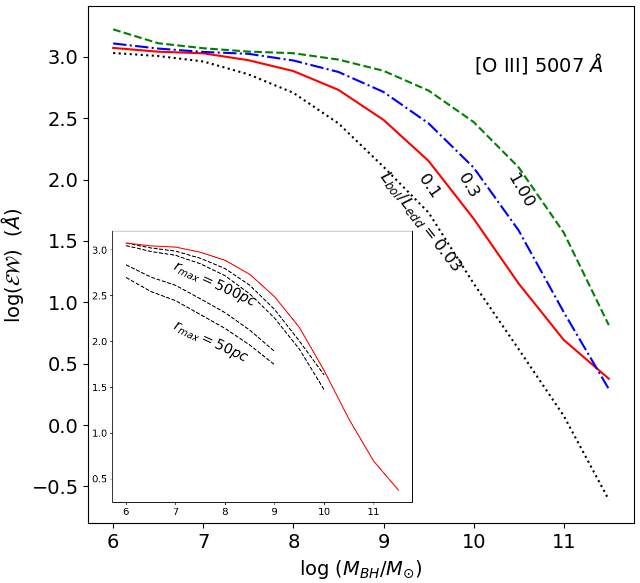}
	\caption{log$\mathcal{EW}$ of [O III] 5007 \AA $ $ as a function of log$M_{BH}/M_{\odot}$ for different $L_{bol}/L_{edd}$. In the inset panel, we have compared, for accretion rate 0.1, modified $\mathcal{EW}$s where the LOC integration was done up to $r = 1000, 500, 100, 50$ pc (dashed black lines, from top to bottom, respectively)}  
\labfig{fig:EW_O3_LOC}
\end{center}
\end{figure}

We want to know if the NELs will have different observable properties as mass
of the BH rises. Hence, we conducted CLOUDY simulations of the NELs, as well.
In the left panel of Figure \ref{fig:NLRContourR}  we show the iso-contours of
calculated EWs (in log) for the [O III] 5700 \AA $ $ line in the log
$n_{H}$-log $r$ plane for $M_{BH}=10^{8.0} M_{\odot}$ and
$L_{bol}/L_{Edd}=0.1$. As mentioned in Section \ref{sec:CLOUDY Calculation}
(Table \ref{tab:NELparameter}), we are using a different range of $\Phi$ and
$n_{H}$ for NELs (as opposed to the right panel of Figure
\ref{fig:BLRContour}). As NLRs are situated further away from the BLRs, the
range of $\Phi(H) = Q(H)/4 \pi r^{2}$ for which NELs are efficiently produced
is lesser than that of BELs \citep{Peterson97}. The EWs are measured relative
to the incident continuum at 4860 \AA $ $ which is then normalised to the
central wavelength of each line. We see that our contour plots match closely
with the contours presented in \citet{Ferguson97} for log$N_{H}$=23. The small
differences are because of the slightly different SED that we use. In the right
panel of Figure \ref{fig:NLRContourR}, we represent the contour for
$M_{BH}=10^{11}M_{\odot}$ to compare the efficiency of line emission at higher
masses. While the highest EW achieved for the $M_{BH}=10^{8}M_{\odot}$ SED is
$10^{3.75}$, that for the $M_{BH}=10^{11}M_{\odot}$ SED is $10^{1.25}$. Thus,
everything else remaining the same, the line becomes fainter by 2.5 dexes. 

Over the entire range of parameter space, we used CLOUDY to calculate the line
strengths for almost all of the lines discussed in \citet{Ferguson97}. However,
for the scope of this paper, we use results of 6 of them, namely H$\beta$ 4861
\AA, [O III] 5007 \AA, [O I] 6300 \AA, H$\alpha$ 6563 \AA, [N II] 6584 \AA $ $
and [S II] 6720 \AA. The individual EWs were then used in Eqs.
\ref{eq:LOCLine3} and \ref{eq:LOCEW1} to calculate $L_{line}$ and
$\mathcal{EW}$. The sensitivity of $\gamma$ and $\beta$ of Eq.
\ref{eq:LOCLine3} in case of NELs are presented in detail in \citet{Ferguson97}
and shows that for dust-free models like ours, $\gamma=-1.25$ and $\beta=-1.4$
fits the observational spectra the best. Hence, we use these indices for our
calculations. In the case of BELs, we had used an upper limit for $r (=
R_{keplerian})$ corresponding to $v_{min} = 1000$ km/s. To make a clear
distinction between BLR and NLR, we use a similar idea, but this time for lower
limit of $r (= R_{keplerian})$ corresponding to $v_{max} = 500$ km/s. The
results are shown in \fig{fig:EW_O3_LOC} (main panel). Note that unlike some of
the BELs (O VI 1036 \AA, CIV 1549 \AA $ $ and He 1640 \AA), the EW of [O III]
5007 \AA $ $ line decreases monotonically with the increase in the mass of the
black hole. $\mathcal{EW}$ of [O III] 5007 \AA $ $ drops by $\sim$ 1.5 dex from
$M_{BH}=10^{8}M_{\odot}$ to $10^{10}M_{\odot}$. This amplitude is similar to
those seen for most of the BELs. The same behaviour holds for all 6 NELs that we
are discussing in this paper.  

To understand the evolution of the [O III] 5007 \AA $ $ line $\mathcal{EW}$, we
again refer to Figure \ref{fig:ionization}. The ionization potential of O II is
similar to that of He I, but the corresponding [O III] line does not exhibit a
turnover in EW, similar to that of He II 1640 \AA $ $ broad emission line. The
EW profile is rather similar to that of the Hydrogen lines H$\alpha$ and
H$\beta$, indicating that it is the shape of the SED at the energy range of the
[O III] line's central wavelength, that determines the strength of the line.
Because [O III] 5007 \AA $ $, is a forbidden line, it is the energy (of the
illuminating SED) at the central wavelength of the line, that is of more
importance than the energy at the IP of O II, as explained by the Stoy
energy balance temperature indicator \citep[see section 5.10
of][]{Osterbrock06}. The evolution of the $\mathcal{EW}$ of all the other 5
NELs would be similar to that of [O III] 5007 \AA $ $, because they too are
forbidden lines.

\begin{figure*}
\begin{center}
\includegraphics[width=0.9\textwidth, trim= 0 10 0 0]{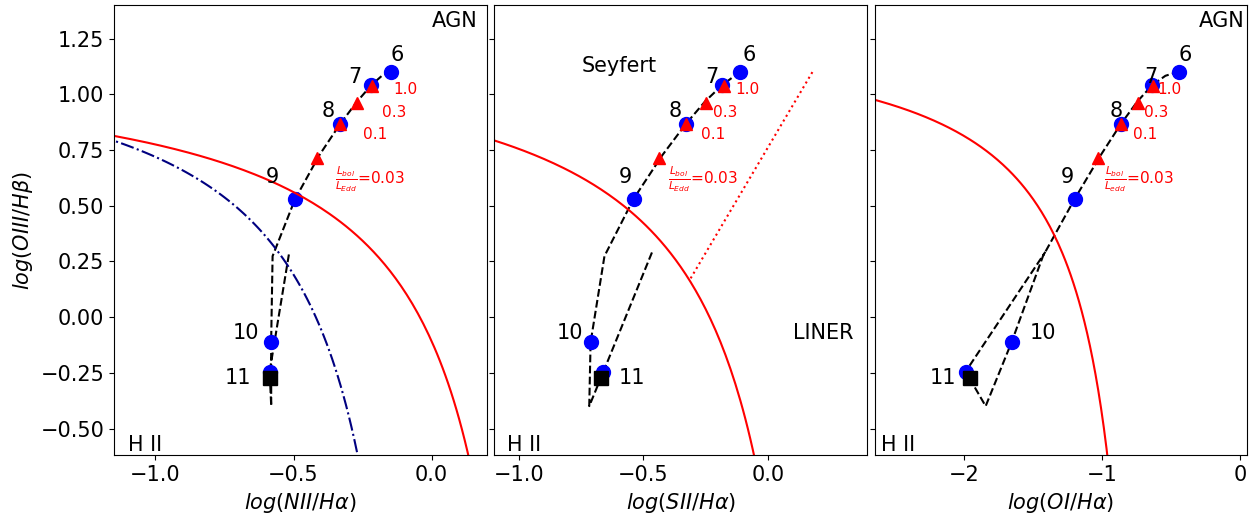}
	\caption{Line luminosity ratios of few observable prominent NELs as a function of BH mass for $L_{bol}/L_{Edd}=0.1$ is plotted on Kewley et al. (2001 \& 2006) plots (red line) and is marked in blue circles and are labeled with log$(M_{BH}/M_{\odot})$ adjacent to them. Line luminosity ratios as a function of accretion rates, but only for $M_{BH}=10^{8}M_{\odot}$ is marked in red triangles. We have plotted a Black square corresponding to the AGN whose total luminosity is $10^{47} erg$ $ cm^{-2} s^{-1}$ to indicate the position of brightest known quasars on our plot.}  
\label{fig:NELLineratios}
\end{center}
\end{figure*}

The $\mathcal{EW}$s of O III 5007 \AA $ $ $\mathcal{EW}$s are higher from that
usually observed for the O III line ($\lesssim 100$ - Figure 1 of
\citet{Risaliti11} shows a distribution; also see \citet{Shen11}). One reason
for the discrepancy is because we did not put any constraint on the upper limit
of $r$ in the LOC integration of NELs. We demonstrate the effect of
incorporating this constrain in the inset of \fig{fig:EW_O3_LOC}.  notice the
comparison, with the solid red curve, of the modified $\mathcal{EW}$s as the
upper limit on $r$ is varied from 1000 pc to 50 pc, through 500 and 100 pc. For
the $r_{max} = 100 pc$ profile, we see that for the $10^8 M_{\odot}$ BH,
$\log(\mathcal{EW}) \sim 2.5$, which yields a value about 0.5 - 1 dex higher
from the predictions of \citet{Risaliti11}. While we understand that the sample
in \citet{Risaliti11} can have BHs with higher mass and/or lower accretion
rate, the second reason for the discrepancy is the use of the uniform/constant
covering factor of 0.24, a value that is motivated by the study of broad lines
and satisfies the BEL $\mathcal{EW}$s, as we have seen in Section
\ref{sec:Broad Emission Lines}. \citet{Baskin05} shows that this value for
covering factor for [O III] 5007 \AA $ $ line ranges between 0.02-0.2. If we
account for this additional factor of 10 (drop in $\mathcal{EW}$), we reach a
good agreement with the observed values reported in \citet{Risaliti11}.  That
we rely on $r_{max}$ to be 100 pc, or lower, is an indication, that we usually
observe more compact NLRs. This initial analysis shows here, that there is
scope of discerning the nature of NLRs (and also BLRs) if a rigorous,
systematic analysis of the variations of the LOC parameters are conducted for
NLRs, a study beyond the scope of this paper.


\section{Line Ratios}

\subsection{Narrow Lines and BPT diagrams}
\label{subsec:Nlr_Bpt} 

Unlike BELs, NELs are not direct signatures of AGN activity. For example, star-forming regions also emit Narrow (of the order of 100 km s$^{-1}$) emission lines. So, making the distinction between the two becomes very important. That is where the BPT diagrams become very useful. BPT diagrams were first plotted by \citet{Baldwin81} to classify galaxies based on their excitation mechanisms. They found an empirical method of separating AGN from star-forming regions (SRFs) based on line ratios.

\citet{Kewley01} used theoretical pure star photoionization models to give a maximum starburst line for some of the important BPT line ratios to distinguish AGN from SFRs more cleanly. We use the same line ratio planes as \citet{Kewley01}. The maximum starburst lines given by \citet{Kewley01} are plotted as the solid red lines in each panel of the Figure \ref{fig:NELLineratios}.

The equations for these lines were as follows:
\begin{equation}
{\textnormal{log([O III]/H$\beta$) > 0.61/[log([N II]/H$\alpha$) - 0.47] + 1.19 }}
\end{equation}
\begin{equation}
{\textnormal{log([O III]/H$\beta$) > 0.72/[log([S II]/H$\alpha$) - 0.32] + 1.30}}
\end{equation}
\begin{equation}
{\textnormal{log([O III]/H$\beta$) > 0.73/[log([O I]/H$\alpha$) + 0.59] + 1.33}}
\end{equation}

There have been other schemes of demarcation as well. For example, the semi-empirical line given by \citet{Kauffmann03} is marked as a dotted-and-dashed blue line in the left panel of Figure \ref{fig:NELLineratios}. As predicted by \citet{Kewley01}, the region below the solid red lines is expected to be the part of the plane populated by SFR line ratios, while the region above represents the AGN activity. As such, these line ratios have also become a method to look for AGN. Our theoretical predictions as a function of $M_{BH}$ in the line ratio planes are traced by the dashed black lines in Figure \ref{fig:NELLineratios}, for $L_{bol}/L_{edd} = 0.1$. We see that for the higher mass black holes ($10^{8.5} M_{\odot}  \lesssim M_{BH} \lesssim 10^{10.75} M_{\odot}$), the line ratios of NELs move into the star-forming regions of \citet{Kewley01} plots. Hence these higher mass black holes will be incorrectly detected as SFR using these diagrams. We have further plotted the line luminosity ratios for different accretion rates $L_{bol}/L_{edd}$  (but only, for $M_{BH}=10^{8}M_{\odot}$) with red triangles in Figure \ref{fig:NELLineratios}. Note that, the less luminous BHs (even for lesser mass $\sim1 0^{8} M_{\odot}$) also start to seep into the SFR. This change is due to the disc SED being cooler, comparable to O-star temperatures, and so produce similar emission lines to H II regions. \textit{Therefore, by using this method to distinguish AGNs from SFRs, we are losing low accretion rate and high mass black holes.} 

\subsection{Line Ratios for BELs}
\label{subsec:BelLR}

\begin{figure*}
\begin{center}
\includegraphics[width=0.9\textwidth, trim= 0 10 0 10]{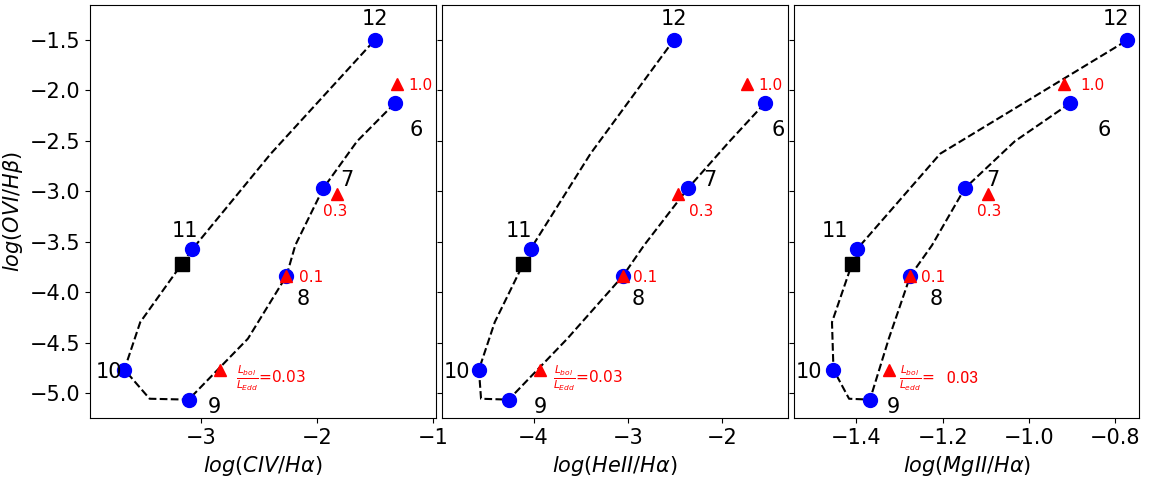}
	\caption{Line luminosity ratios of few observable prominent BELs as a function of BH mass for $L_{bol}/L_{Edd}=0.1$ is marked in blue circles and are labeled with log$(M_{BH}/M_{\odot})$ adjacent to them. Note that we use the line 1640 \AA $ $ He II emission line in the middle panel, here, and in all subsequent figures of line ratios of BELs. We see that the degeneracy is lost when we plot the ratios on multi-line plane. Line luminosity ratios as a function of accretion rates, but only for $M_{BH}=10^{8}M_{\odot}$ is marked in red triangles. We have plotted a Black square corresponding to the AGN whose total luminosity is $10^{47} erg$ $cm^{-2} s^{-1}$ to indicate the position of brightest known quasars on our plot.}  
\label{fig:BELLineratios}
\end{center}
\end{figure*}

\begin{figure}
\begin{center}
\includegraphics[width=0.46\textwidth, trim= 0 10 0 10]{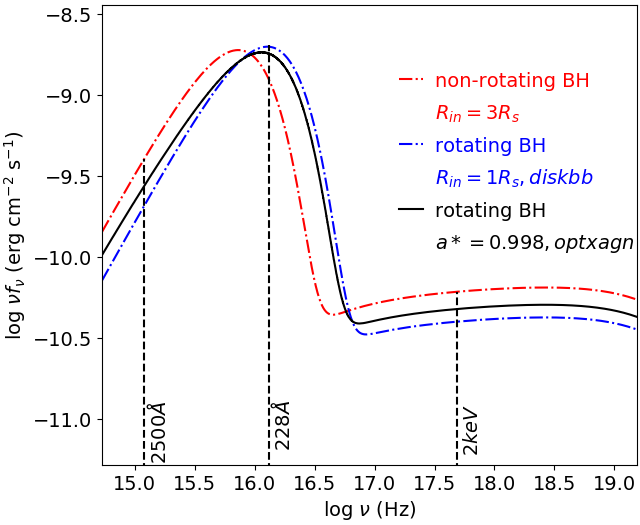}
	\caption{Comparison of SEDs of maximally spinning and non-spinning BH for $M_{BH}=10^{8}M_{\odot}$. Note that the peak of $diskbb$ is more energetic and power-law component is less prominent for spinning BHs.}
\label{fig:SED_spin}
\end{center}
\end{figure}

\begin{figure*}
\begin{center}
\includegraphics[width=\textwidth, trim= 0 20 0 10]{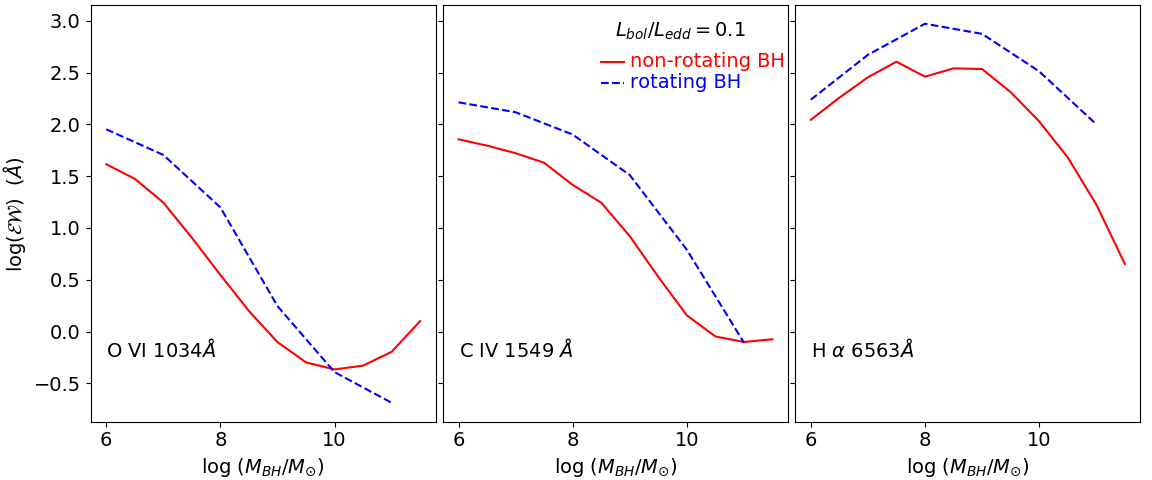}
	\caption{Evolution of some of the $\mathcal{EW}$ as an extension from Figure \ref{fig:BEL_EW}. Even though the $\mathcal{EW}$ of spinning BHs are stronger than the non-spinning BHs, the trend remains.}
\label{fig:EW_spin}
\end{center}
\end{figure*}

\begin{figure}
\begin{center}
\includegraphics[width=0.45\textwidth, trim= 0 10 0 10]{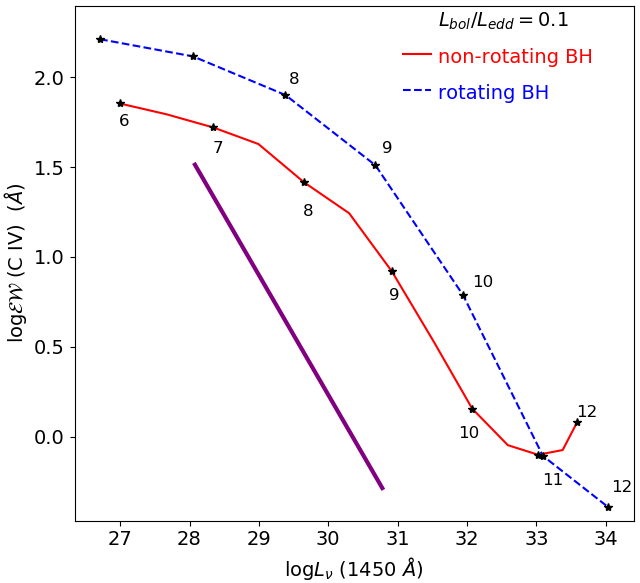}
	\caption{The Baldwin Effect demonstrated for the zero spin and highly spinning black hole. }
\labfig{fig:Baldwin_spin}
\end{center}
\end{figure}

As seen in Figure \ref{fig:BEL_EW}, for some emission lines a particular log$(\mathcal{E} \mathcal{W})$ corresponds to two different BH masses. O VI 1034 \AA, C IV 1549 \AA $ $ and to some extent He II  1640 \AA show degeneracy at the high mass end whereas Mg II 2798 \AA, H$\beta$ 4861 \AA and H$\alpha$ 6563 \AA $ $ show degeneracy at the low mass end. Therefore, if we try to estimate the black hole mass using just the observed line strengths, we might end up with two very different values for BH masses agreeing to the observed line strengths.  To remove this degeneracy we plot the line luminosity ratios on a multi-line plane as a counterpart to the BPT diagrams for NELs (Figure \ref{fig:BELLineratios}). Blue circles in the Figure \ref{fig:BELLineratios} are line luminosity ratios for different masses for accretion rate $L_{bol}/L_{edd}=0.1$ traced by a black dashed line. The red triangles are line luminosity ratios for different accretion rates $L_{bol}/L_{edd}$, but only for $M_{BH}=10^{8}M_{\odot}$. Such plots will be of use to determine the black hole mass just by using the observed line ratios. This may be a new tool for selecting HMBHs.

\section{Spinning Black holes}
\label{sec:rotatingBH}

\begin{figure*}
\begin{center}
	\hspace{1 mm}\includegraphics[width=0.9\textwidth, trim= 0 0 0 0]{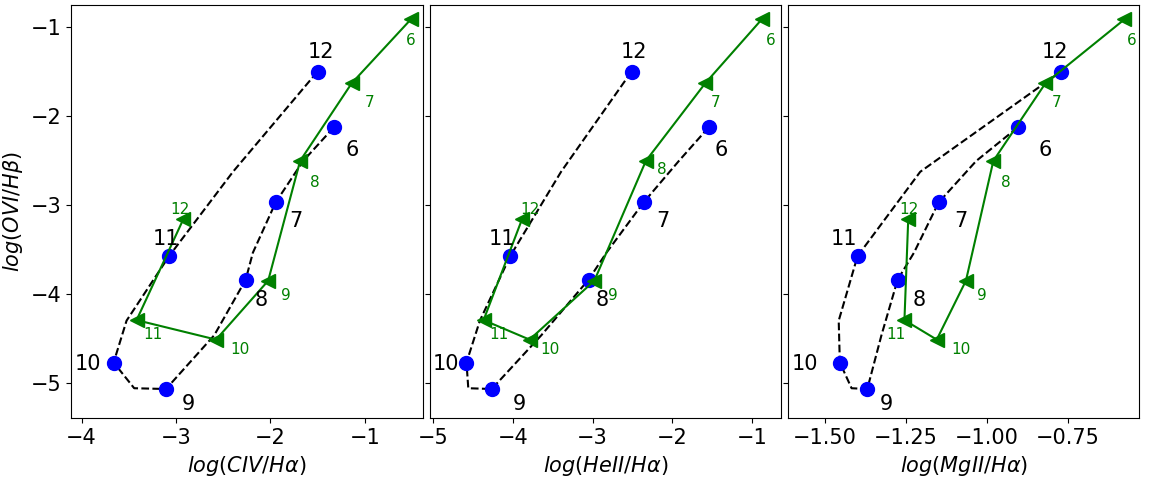}
\includegraphics[width=0.9\textwidth, trim= 0 10 5 0]{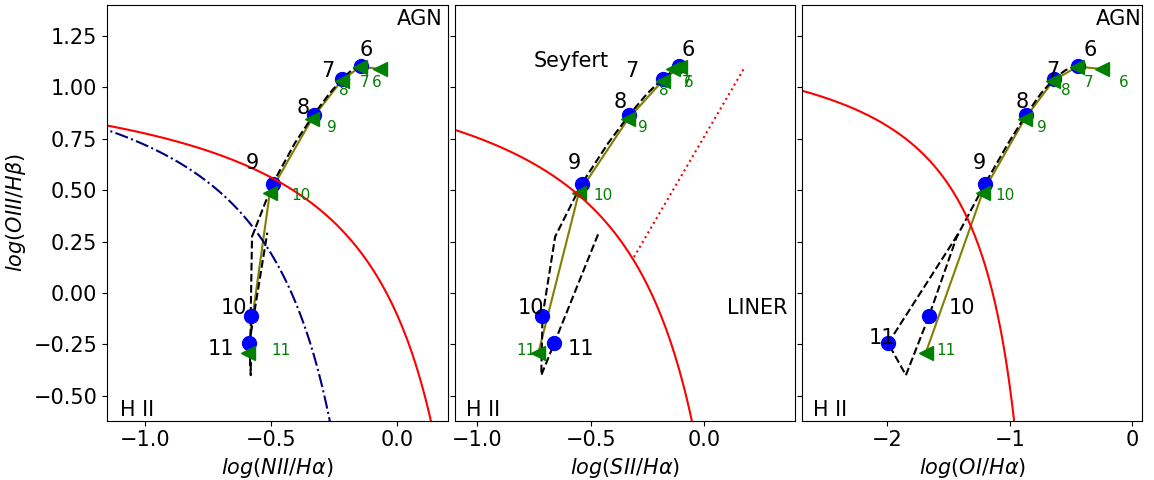}
	\caption{Extending Figures \ref{fig:BELLineratios} and \ref{fig:NELLineratios}, we compare the broad (top panels) and narrow (bottom panels) emission line ratios of spinning BHs. Green triangles joined by the solid green line, in the figure, correspond to spinning BHs and are labeled with log($M_{BH}/M_{\odot}$) adjacent to them (smaller green fonts). For comparison, we retain the line ratio profiles for the non-spinning black holes through the black dashed curves joining the blue circles. }
\label{fig:Lineratios_spin}
\end{center}
\end{figure*}


BHs are known to have spin associated with them \citep{Reynolds14,
Reynolds19}. Our analysis in the previous sections was motivated to look at the
influence of mass variation on the changes in the line emission strengths. Just
as the change in mass varies the SED, so does the variation in the spin of the
BH (see \citet{Bertemes16} for a detailed discussion). Hence, in this section,
we look into the effect of the spin of the BHs. We do the same rigorous
analysis of emission line strengths, as in previous sections, but comparing,
only, the `maximally spinning' BHs to the zero spin ones (analysed in the
previous sections), because they will show the maximum deviation from the spin
zero case. Results for any intermediate spin is beyond the scope of this paper
- it is expected that results for the intermediate spin BHs will simply, lie in
between. The more quantitative rigorous analysis will be conducted in future
publications where we intend to compare theoretical predictions with
observations. 

$Diskbb$ is a SED package that does not have spin as an explicit input
parameter. However, since all SED generation in the previous sections relied on
$diskbb$, we find a way to tweak and use the same package, for BHs with spin,
for the sake of uniformity. We know that the innermost stable circular orbit of
the accretion disc of a spinning black hole moves relatively closer (compared
to that of a non-spinning or less spinning one) to the BH. Thus, in Equations
\ref{eq:Temperature} and \ref{eq:Adbb}, if we opt for a smaller value of
$R_{in}$, while correctly adjusting for $\eta$, then we can mimic the SED of a
spinning BH, even while using the package $diskbb$. To understand what will be
a reasonable $R_{in}$ modification to assume, we compare with an SED generated
by the package $optxagn$ \citep{Done12}, from Xspec version 12.10. In $optxagn$
we generated the accretion disc SED of a maximally spinning ($a^{\star} =
0.998$) BH of mass $10^8 M_{\odot}$, accreting at $L_{disk} = 0.1 L_{edd}$.
Keeping the mass and the accretion rate constant for the BH, when we used
$diskbb$ input parameters, we found a very good match of the SED, for $R_{in} =
R_s = 2R_G$ (resulting in $\eta = 0.75$), while comparing $L_{\nu} (2500 \AA)$
and the energy where the SED peaks (two important aspects relevant to the
analysis in this paper). Hence, we proceed to make this a norm for SEDs in this
section (and the paper) for BHs with spin - we use $diskbb$, with $R_{in}$ set
to $R_s = 2R_G$. 
We understand that for a non-spinning black hole, $R_{in} = R_s$ (or $\eta
= 0.75$) is less than the allowed limit of $3R_s$. Thus, for $diskbb$, which is
a package for spin zero BHs, using $R_{in} = R_s$ is unphysical - we made this
choice as an artificial fix in the package which cannot include the effects of
spin, in any other way. Figure
\ref{fig:SED_spin} shows the comparisons of the resultant SEDs for $10^8
M_{\odot}$, accreting at $L_{bol} = 0.1 L_{edd}$ ($\theta = 30$ degrees).
Compared to the Schwarzschild SED (red, dotted-and-dashed line), the peak of
the SED for the maximally spinning BH is shifted by 0.25 dex (and 0.15 dex),
higher in frequency, for the $diskbb$ (and the $optxagn$) SED. At the HeII
ionization edge (at 228 \AA), the maximally spinning $diskbb$ (and $optxagn$)
SED is 0.2 dex (and 0.16 dex) higher in normalization than the Schwarzschild
SED. We generate SEDs corresponding to a range of $10^6 \leq M_{BH} \leq
10^{11}$, but, this time, with an interval of 1.0 dex. These SEDs are then used
through CLOUDY to predict emission line strengths for both BELs and NELs, in
the same way that have been discussed in the previous sections, for the
Schwarzschild BHs. The results of this analysis is presented in Figures
\ref{fig:EW_spin} through \ref{fig:Lineratios_spin}. 

We see that in the case of spinning BHs (of same mass $M_{BH} = 10^8
M_{\odot}$), $\mathcal{EW}$ for C IV 1549 \AA $ $ is higher (by 0.45 dex) than
that for the Schwarzchild BHs and this factor decreases for HMBHs (Figure
\ref{fig:EW_spin}). However, the overall trend of declining $\mathcal{EW}$, for
all the lines, as a function of mass, still remains true even in the extreme
case of maximum spin. We had noted earlier that for the O VI and C IV lines,
the $\mathcal{EW}$ turns over at the high mass end, for Schwarzchild BHs with
$L_{bol}/L_{edd} \leq 0.1$. However, we do not see such turn-over for the
spinning BHs. This is not surprising, because, for the same mass, the SED of
the spinning BH has the same effect as the SED of BH with a higher accretion rate
(see Figure 6). On the other hand, the $\mathcal{EW}$ for H $\alpha$ drops much
less, for spinning BHs!

\fig{fig:Baldwin_spin} demonstrates that the Baldwin effect, for the C IV line
holds fort, for spinning BHs; in fact, higher the spin, better would be the
match of the slope of the profile with the value -2/3 (-0.66), since the
Baldwin effect is $\mathcal{EW} \sim L^{-2/3}_{\nu}(1450 \AA)$. For the
non-spinning black hole, the slope is -0.53 for the mass range $10^8 - 10^{11}
M_{\odot}$ and for the spinning black hole, the slope is -0.67, for the mass
range $10^9 - 10^{12}$.

Figure \ref{fig:Lineratios_spin} shows the line ratios for the BELs (top
panels) and NELs (bottom panels). There are clearly quantitative differences in
the line ratios, but the overall trend remains the same even for the spinning BHs.
Even though higher mass spinning BHs fall into H II region of Kewley plots, it
happens only for $ M_{BH} \geq 10^{10}M_{\odot}$ (as opposed to $ M_{BH} \geq
10^{9}M_{\odot}$ in the case of non-spinning BHs).


\section{Emission Line strengths in lower mass Schwarzchild black holes}
\label{sec:lowNLR}

\begin{figure*}
\begin{center}
\includegraphics[width=\textwidth, trim= 0 25 0 10]{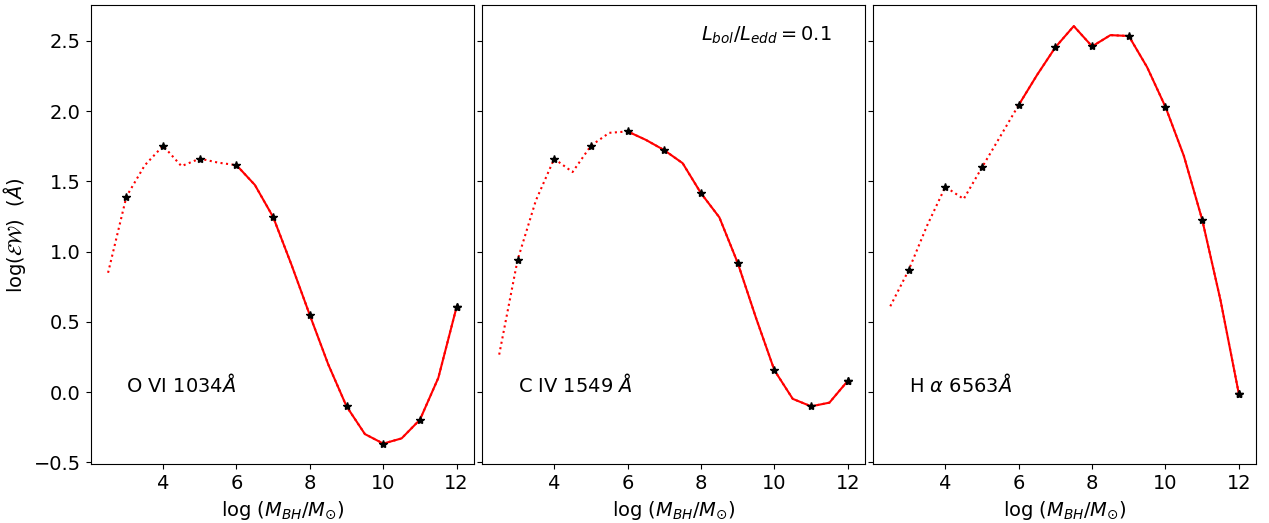}
	\caption{Evolution of some of the $\mathcal{EW}$ as an extension from Figure \ref{fig:BEL_EW}. The $\mathcal{EW}$ for the lower mass black holes (traced by red dotted lines joining the stars) are added to show the variation across the entire BH mass range.}
\labfig{fig:BELs_EW_lowM}
\end{center}
\end{figure*}

We have mainly focused in this paper on the super to hyper massive black
holes, with the motivation to see how emission line strengths evolve as this
transition is made. In \citet{Chakravorty14}, on the other hand, the authors
had concentrated their efforts on the low mass black holes. Since, accretion
disks of lower mass ($\leq 10^6 M_{\odot}$) peak at extreme ultraviolet, the
part of the SED which ionises the line emitting gas, is entirely dominated by
the disk SED. Hence, in \citet{Chakravorty14} had ignored the power-law
component. Reconstructing a disk plus power-law SED for the low mass black holes
would require involved analysis of observations, and such an exercise is beyond
the scope of this paper. However, following the cue of \citet{Chakravorty14},
we can rely on `approximate' SEDs for the lower mass Schwarzchild black holes,
where the disk is the only component. After construction of these SEDs, for
$L_{bol}/L_{edd} = 0.1$, and $10^2 \leq M_{BH}/M_{\odot} \leq 10^5$ (at steps
of 1.0 dex) we run them through CLOUDY to calculate the emission line
strengths. Following the same methods, on the line emissions, as for the higher
mass BHs, we try to build a `continuum' of emission line properties across the
entire mass range $10^2 \leq M_{BH}/M_{\odot} \leq 10^{12}$. While, the SEDs
for the lower masses, may not be accurate, we get a few interesting insights.

\fig{fig:BELs_EW_lowM} shows that there is a `sweet spot' in BH mass range
$10^6 - 10^9 M_{\odot}$, where the strong emission lines like C IV and
H$\alpha$ peak and their $\mathcal{EW}$ drop sharply, on both sides of this
mass range! This is a striking result, because this gives us an insight into
the resultant selection bias that all emission line surveys must have in favour
of $10^6 - 10^9 M_{\odot}$ BHs. This is in fact where most AGN BH masses are
found \citep[e.g. in][and references therein]{Shen11}.  Given this bias, it
could be that the true range of BH masses is much broader than has been
supposed. O VI seems to have a constant strength down to masses, as low as
$10^4 M_{\odot}$. So, in the search for intermediate mass BHs, this would be
the best emission line to target.  OVI is hard to detect, however, as it
requires far UV spectroscopy at low redshifts, as from FUSE \citep{Kriss04},
while at high redshifts it is almost always absorbed by the Lyman alpha forest.

\begin{figure*}
\begin{center}
	\hspace{1 mm}\includegraphics[width=0.9\textwidth, trim= 0 0 0 0]{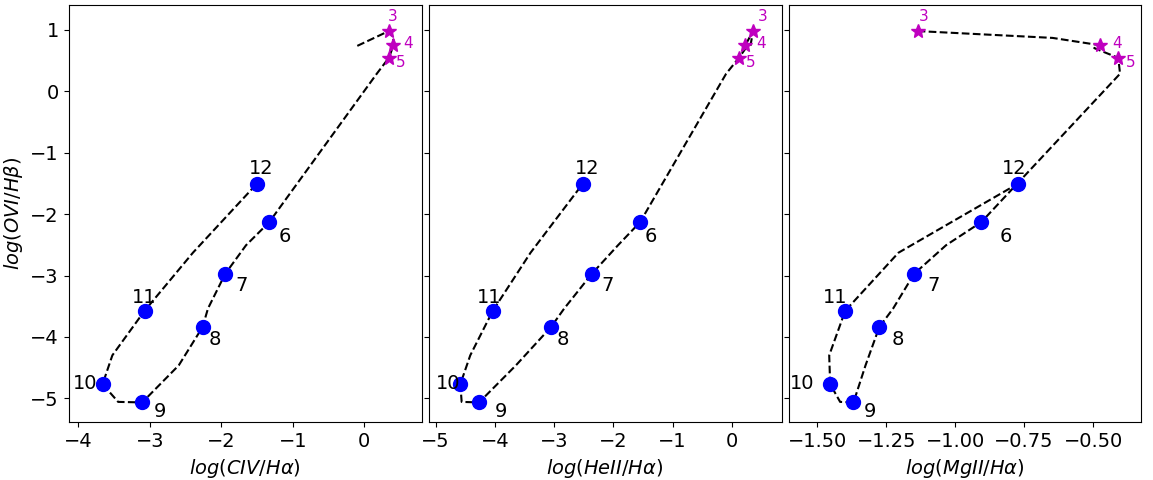}
\includegraphics[width=0.9\textwidth, trim= 0 10 5 0]{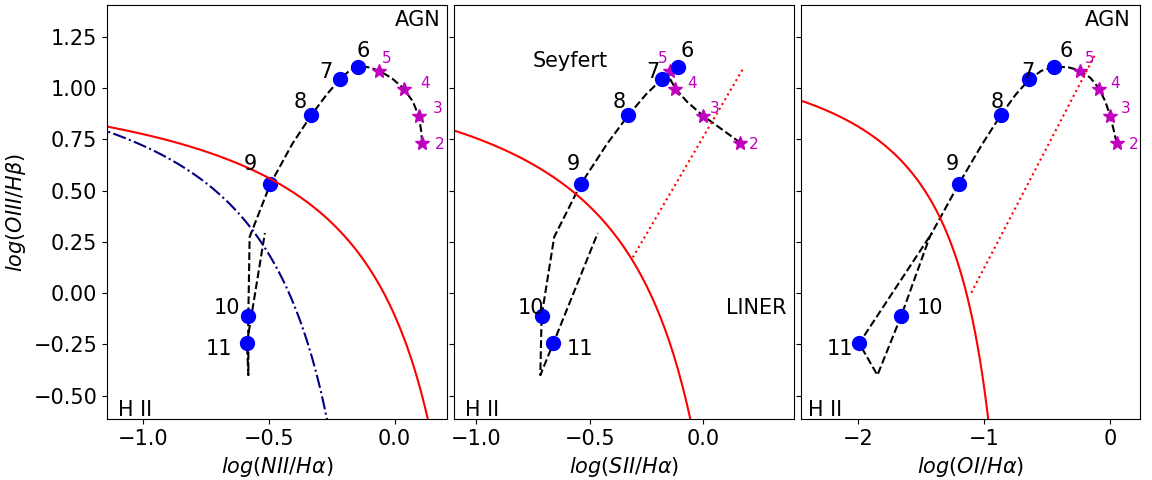}
	\caption{As an extension from Figure \ref{fig:NELLineratios}, we have calculated the line ratios for lower mass black holes, $10^{2} M_{\odot}<M_{BH}<10^{6}M_{\odot}$ (extended black dashed line, studded with magenta star symbols) at a steps of 1 dex.  We have used just the $diskbb$ component of the SED. }
\label{fig:Lineratios_low}
\end{center}
\end{figure*}

Figure \ref{fig:Lineratios_low} shows the line ratios for the entire mass range, for
BELs (top panels) and NELs (bottom panels). The BPT diagrams of the ratios of
the different narrow emission lines is a common tool, now, not only to
distinguish AGN from start forming regions but even to distinguish Seyfert-like
AGN from LINER-like AGN. Note that \citet{Chakravorty14} had presented some
results about how the $\mathcal{EW}$ of broad emission lines will look like,
albeit with much less rigour than has been used in this paper. However, there
has been no investigation of the evolution of the BPT diagrams, concentrating
on the low mass end. In \citet{Kewley06}, the authors had introduced a dividing
line between Seyferts and LINERS, shown as the dotted red line on the (middle)
log([O III]/H$\beta$) vs log([S II]/H$\alpha$) and the (right) log([O
III]/H$\beta$) vs log([O I]/H$\alpha$) line ratio planes.
In Figure \ref{fig:Lineratios_low}, bottom panels, while in the middle panel even the low
mass BHs are consistent with Seyfert-like line ratios (except for
log($M_{BH}/M_{\odot}) \lesssim 2$), in the right panel, BH with masses lower
than $10^{4.5} M_{\odot}$ are consistent with being LINERS.  

A more rigorous calculation about the low mass BH NELs is beyond the scope of
this paper, because this paper deals with HMBHs.

\section{Discussion}
\label{sec:Discussion}

\subsection{On Truncated discs and Radiatively Inefficient Accretion flows}
\label{subsubsec:Riafs}

To model the radiation from the accretion disc, we have used the `thin disc
paradigm' and used the Shakura Sunyaev model \citep{Shakura73,Frank02}
where L$\sim \dot{m}$, however, when the accretion rate
decreases below $\dot{m} < 0.01$, L$\sim \dot{m}^2$. The accretion flow close
to the BH becomes very different, the disc becomes optically thin and
geometrically thick and radiatively inefficient. We are pursuing the
calculations for such radiatively inefficient accretion flow (RIAF) in a
separate analysis. In such cases, there are often signatures that the innermost parts of the accretion disc are truncated, where the radiatively
inefficient flow dominates, and there is an outer thin disc. SEDs,
corresponding to such scenarios are different flavours of the may be similar to
the RIAF SEDs. We will address such SEDs, also, in the RIAF paper. Since these
SEDs are predicted to be `dim' in UV and optical but bright in X-rays, in
addition to looking at the emission line properties, we will also look into the
properties of the X-ray warm absorbers for SMBH harbouring RIAFs.

\subsection{ On the ~ 150 eV soft excess component of AGN SEDs}
\label{subsec:SoftExcess}

X-ray observations from ROSAT and XMM-Newton show the presence of unaccounted
excess intensity at $E < 1 \kev$ known as the {\it soft excess} component, in
the observes spectra of type1 AGN. \citep{Elvis85, Brinkmann92, Buehler95,
Pounds02}. This component is usually by a blackbody with temperature $T_{se}
\sim 100 - 200 \ev$ (i.e. peaking at $\sim 282 - 564 \ev$) \citep[][and
references therein]{Matsumoto04, Piconcelli05, Porquet04, Vignali04}. The ratio
of the {\it soft excess} luminosity to power-law luminosity (between 0.1 - 10
keV) is independent for each object, for e.g. it is 0.04 in Mkn 304
\citep{Piconcelli05} and $\gtrsim 1.0$ in Ark 564 \citep{Vignali04}. The
$diskbb$ model, though represents UV SED of the observed AGN spectra
satisfactorily, cannot model the {\it soft excess} component because centres of
SMBHs are too cold to reach the peak temperature at  $\sim 0.5 \kev$. {\it soft
excess} component can be explained only by making modifications to the $diskbb$
model \citep{Czerny87, Korista97b} or by introducing an additional spectral
component.

In this paper, we have ignored the {\it soft excess} as a component of the AGN
SED. Because the physical origin of the {\it soft excess} component is not
certain, hence it is not easy to relate it, in a systematic way, to either the
accretion disc component or the power-law. There is no empirical relation that
will guide us to find the evolution of the {\it soft excess} component as a
function of the BH mass. The task of finding the correct prescription to relate
these components is beyond the scope of this paper. In our future attempts at
deriving more physically motivated SED, we can use (for example) optxagn to
generate the {\it soft excess} as a function of the disc component (which will
have a dependence on the mass of the BH).

The {\it soft excess} component could be further ignored, for this paper,
because of the energy range where it dominates. As mentioned above, {\it soft
excess} peaks at $\sim 282 - 564 \ev$. The ions emitting the broad and narrow
emission lines, discussed in this paper, are unlikely to be influenced by
photons with such high energies - for example, the IP of O V is $\sim 114$ eV
and the central wavelengths of all the lines are in the UV range. So, for the
purpose of this paper, the {\it soft excess} would have been a redundant
component.   

\section{Conclusions}

\begin{itemize}
	\item The motivation of this paper was to test, for Schwarzchild BHs, a) if there is a higher mass cut-off above which black holes cannot efficiently produce typical BELs such as O VI 1034 \AA $ $, C IV 1549 \AA $ $, He II 1640 \AA, Mg II 2798 \AA, H$\beta$ 4861 \AA $ $ and H$\alpha$ 6563 \AA $ $ used to detect AGNs, b) how would ratios	of the NELs evolve for the Hypermassive ($\gtrsim 10^9
		M_{\odot}$) black holes, in the standard BPT diagrams involving H$\beta$ 4861 \AA, [O III] 5007 \AA, [O I] 6300 \AA, H$\alpha$ 6563 \AA, [N II] 6584 \AA $ $ and [S II] 6720 \AA.  
	\item We wanted to probe the Hypermassive Black Holes (HMBHs) with mass	$\geq 10^8 M_{\odot}$. The first step in the analysis was to build a method of producing systematic mass dependent AGN SEDs for the range $10^6 - 10^{12} M_{\odot}$. For HMBHs, the peak of the disc component moves to lower	energies and hence the important conventional wavelength of	2500 \AA $ $ sees a different part of the disc radiation in case of the most massive HMBHs. We used the latest empirical results known, for the Supermassive Black Holes and used judicious extrapolations to form a suite of SEDs for all BHs in the aforementioned mass range.  
	\item These suites of SEDs were used in CLOUDY to calculate emission line strength for gas clouds with a wide range of densities and	placed over a wide range of distance from the black hole. This	exercise was done for both the (popularly known) broad and	narrow emission lines. The CLOUDY calculated EWs of the individual clouds were then suitably added and averaged using the `Locally Optimally Emitting Cloud' model to predict the final effective equivalent width $\mathcal{EW}$	and luminosity of the emission lines. These emission line strengths and their ratios were then used to predict observables for the HMBHs.  
	\item We found that the optical BELs, like H$\beta$ 4861 \AA $ $ and H$\alpha$ 6563 \AA $ $ are not the best tracers to look for HMBHs because these lines' $\mathcal{EW}$ drop by about a factor of $\sim 100$ as mass increases from $10^8$ to $10^{10}
		M_{\odot}$. This is a very important point to note because large optical surveys like SDSS use these optical broad lines	to look for AGN activity. If we want to look for HMBHs, the	ultraviolet line O VI 1034 \AA $ $ would be the most useful because its strength undergoes a significant and interesting turnover, from their drop, in intensity, for the highest mass	black holes.
	\item The Baldwin effect is clearly reproduced by the changing SEDs in	our calculations. C IV $\mathcal{EW}$ scales as $\sim
		L^{-2/3}_{\nu}(1450 \AA)$ for the super to hypermassive BHs ($\sim 10^8 - 10^{10} M_{\odot}$). It approximately holds for	accretion rate, as low as 0.03, but the relationship matches better for higher accretion rate (or higher spin of the BH), and also spans a larger mass range (up to $\sim 10^{11}
		M_{\odot}$).
		\item Studying the BPT diagrams for the NELs, reveals that BHs with the highest mass with/or low accretion rate may	have line ratios consistent with star-forming regions, thus	making it very difficult to find them in this way. This, further, implies that the number of actively star-forming regions are being overestimated in large survey samples like SDSS and the number of massive black holes are being underestimated. This effect adds to the above-mentioned bias against HMBHs in broad line detection). Such a realisation should have {\bf a profound} effect on the luminosity functions of both these classes of objects, namely AGN and normal (but actively star-forming) galaxies.  
	\item Following the principles of BPT diagrams, we have proposed line ratio diagrams for the BELs, too. The log$(\mathcal{EW})$	profile is often degenerate in mass, which can be broken, by using the line ratio plots for BELs. Note that above mentioned	degeneracy is pronounced and noticeable because we are dealing with a much wider mass range here.
	\item After considering Mass and accretion rates as two main BH parameters which alter the SED of the accretion disc, we also paid attention to the spin of the BH. Since maximally	spinning black holes will have the most different SEDs,	compared to Schwarzschild ones, we compared these two, for the	super to hyper massive BH mass range, while holding the accretion rate constant at 0.1. The effect of the spin turned out to be similar to the effect of the accretion rate, for the BH of same mass. Both of these parameters tend to increase the peak energy $E_{max}$ where the SED has a maxima. The drastic fall off of the line $\mathcal{EW}$ with increasing mass cannot be arrested, for even the spinning black holes, they also satisfy the Baldwin effect. The spinning HMBHs also plunge into the SFR region of the BPT diagram, only for relatively higher masses. Only, if we have a maximally spinning high accretion rate ($\sim 1.0$), HMBHs, then they may be populating the AGN region of the BPT diagrams.   
	\item We have further, investigated the change of $\mathcal{EW}$ for lower masses ($\leq 10^6 M_{\odot}$). \textit{We find that the strong observable emission lines like the C IV and H$\alpha$ favour a mass range $10^6 - 10^9 M_{\odot}$ and drop off sharply, in EW, on both sides of this mass range! This is a striking result, because this gives us an insight into the resultant selection bias that all emission line surveys must	have in favour of $10^6 - 10^9 M_{\odot}$ BHs. This is in fact	where most AGN BH masses are found. Given this bias, it could	be that the true range of BH masses is much broader than has been supposed.} O VI is the best emission line to look for	intermediate mass BHs as it maintains a constant $\mathcal{EW}$ down to $10^4 M_{\odot}$. The BPT diagrams for lower masses show that for some line ratios, they tend to behave like	LINERS.
	\item Our results show that weak emission lines from HMBHs would be	very difficult to detect using the current optical facilities.	However, these same results establish a benchmark which can be used by future 30 m class optical telescopes to look for HMBHs.
		
\end{itemize}


\section{Acknowledgements}
We thank the anonymous reviewer for useful comments and suggestions that helped us improve the quality of this manuscript significantly. HB acknowledges support from IISc, where a major fraction of this work was done during her project. SC was supported by the SERB National Postdoctoral Fellowship (File  No.PDF/2017/000841). NR acknowledges support from the Infosys Foundation through the Infosys Young Investigator grant.

\appendix

\section{Adding disk and power-law without extrapolation relation for HMBHs}
\labsecn{sec:Lusso}

\begin{figure}
\begin{center}
\includegraphics[width=0.45\textwidth, trim= 0 10 0 10]{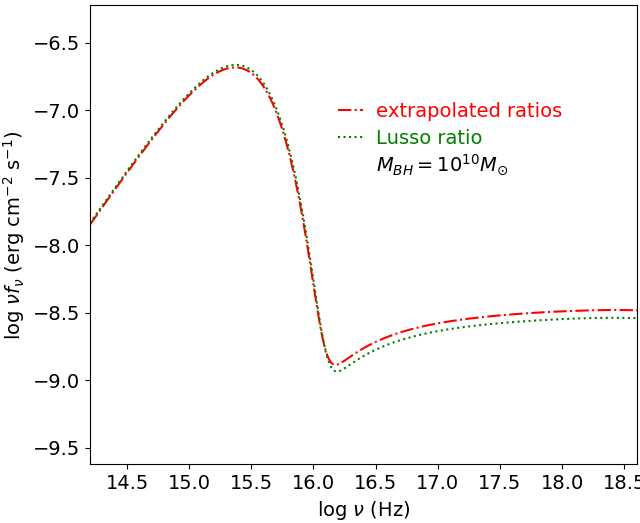}
	\caption{Comparison of SED (called `Lusso ratio') constructed using ``just'' Eqn. 5 (\citet{Lusso16}) vs SED (called `extrapolated ratio') using our more rigorous method using Figure 2 and Eqn. 6, for black hole of mass $10^{10} \, M_{\odot}$ and $L_{bol}/L_{edd} = 0.1$.}
\label{fig:SedLusso}
\end{center}
\end{figure}

\begin{figure*}
\begin{center}
\includegraphics[width=\textwidth, trim= 0 25 0 0]{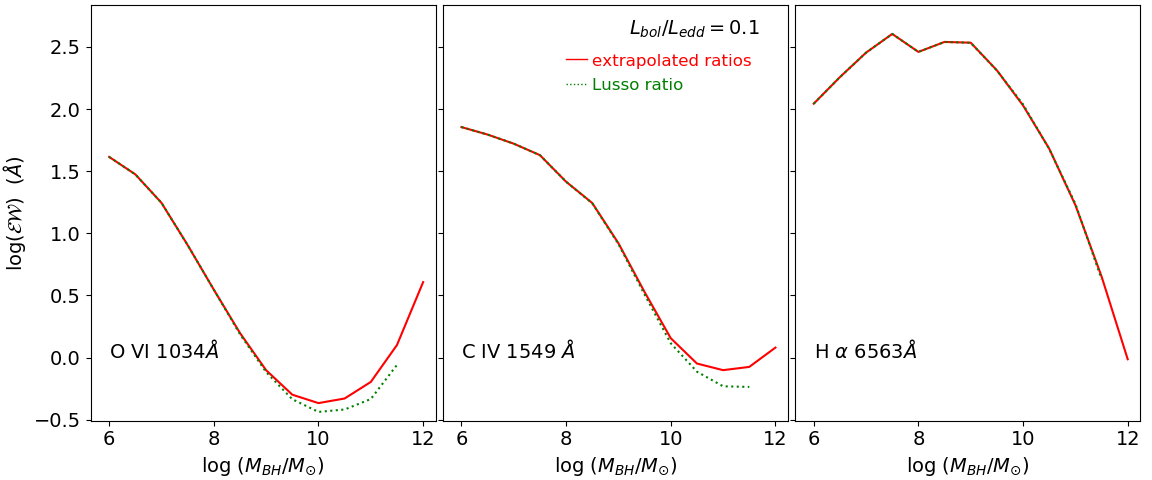}
	\caption{Comparison of some of the $\mathcal{EW}$ as an extension from Figure \ref{fig:BEL_EW}. The SED generation schemes are marked in the Figure. }
\label{fig:EWLusso}
\end{center}
\end{figure*}

In \subsecn{subsec:AgnSed}, we have described that to build our SEDs, we use a
method of extrapolation to scale the $L_{pl}/L_{disk}$ for HMBH, while using
Equation \ref{eq:Lusso&Risaliti} for the supermassive black holes. This is the
method, which has been used throughout the paper and we call this method as
`extrapolated ratio' in this section. One might argue that Equation
\ref{eq:Lusso&Risaliti} does not distinguish between super and hyper massive
black holes and so this equation should be used directly, even for the HMBHs
and we should not be extrapolated. However, these extrapolations, which become
important only for the HMBHs, are based on the physics of accretion discs.
However, in this section, we show that if we directly used Equation
\ref{eq:Lusso&Risaliti} for the highest masses, to scale the power-law with
respect to the disk, and add the two components, the differences in the results
would be minor, and effectively has no difference in the qualitative
inferences. We call this second method of creating the SED (relevant in the
context of the HMBHs, only) as `Lusso ratio', in this section. 

\begin{figure*}
\begin{center}
	\hspace{1mm}\includegraphics[width=0.9\textwidth, trim= 0 0 0 0]{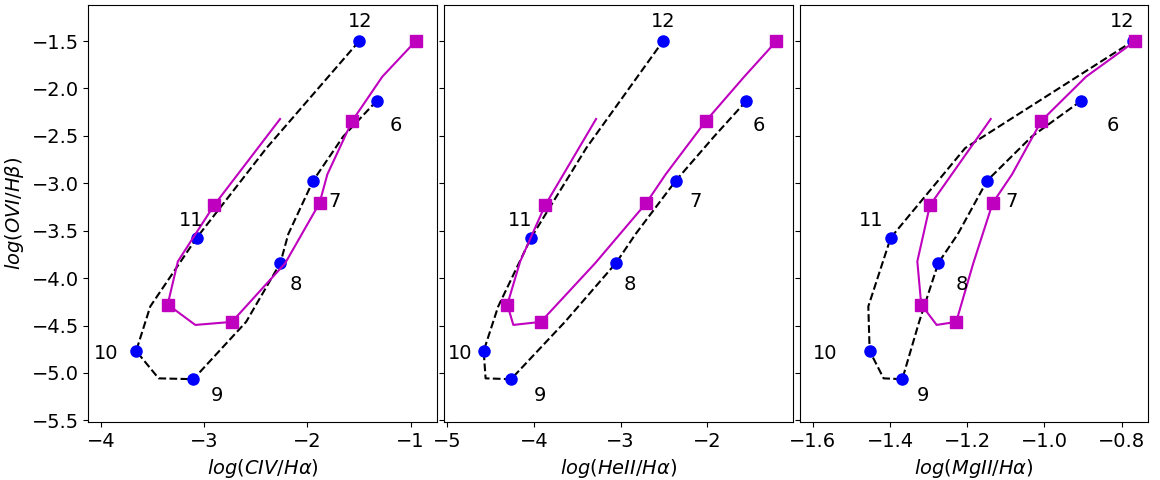}
\includegraphics[width=0.9\textwidth, trim= 0 10 5 0]{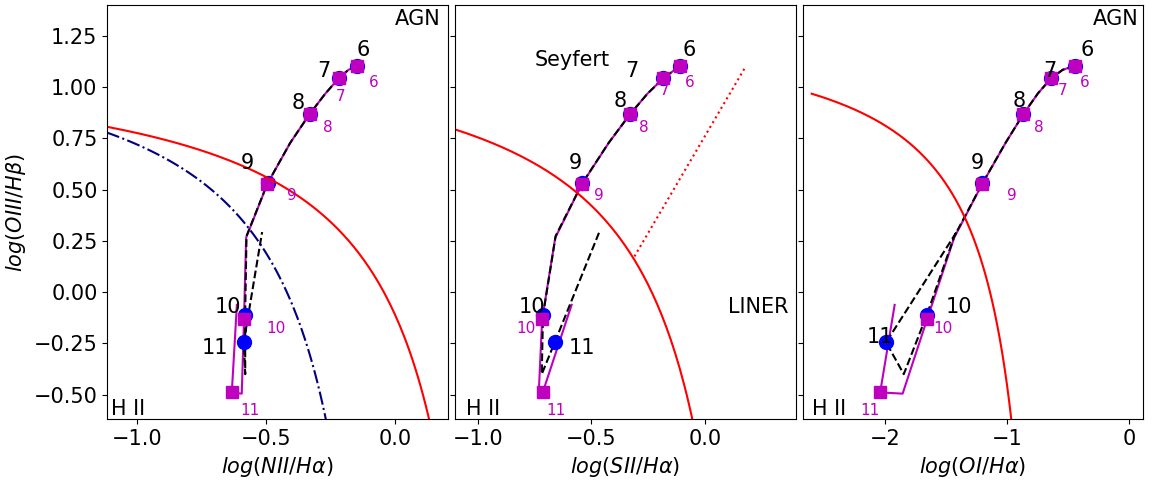}
	\caption{As an extension from Figures \ref{fig:NELLineratios} and \ref{fig:BELLineratios}, we have calculated the line ratios for BELs (top panels) and NELs (bottom panels), to show the comparison of effects of SEDs, in Figure \ref{fig:SedLusso}, but for all the higher masses, for $L_{bol}/L_{edd} = 0.1$. The solid magenta curves represent the results of the `Lusso ratio' method. {\color{red}}}
\label{fig:LineRatiosLusso}
\end{center}
\end{figure*}

\begin{itemize}
	\item In Figure A\ref{fig:SedLusso}, we show the SEDs of BHs of same
		mass $10^{10} M_{\odot}$ and $L_{bol}/L_{edd} = 0.1$, but
		different schemes of SED generation, as labeled. We
		demonstrate that the variation in SED is minor. Such SEDs were
		generated for $L_{bol}/L_{edd} = 0.1$, but only for BHs whose
		SEDs were generated using `extrapolated ratio' method, namely
		the mass range $> 10^{8.25} M_{\odot}$. Note that BHs with
		lower mass, the `Lusso ratio' method was anyway, used. So
		whatever differences arise because of the variation of scheme
		of SED creation, will affect the results for only the higher
		mass BHs. 
	\item Referring to Figures A\ref{fig:EWLusso} and
		A\ref{fig:LineRatiosLusso} we see that there is absolutely no
		qualitative difference to the results. 
\end{itemize}


\end{document}